\documentclass{aa}  

\usepackage{hyperref}
\usepackage{graphicx}
\usepackage{tabularx}
\usepackage{xcolor}
\usepackage{txfonts}
\usepackage{xspace}
\definecolor{cobalt}{rgb}{0.0, 0.28, 0.67}

\def\HI{\ion{H}{I}}

\newcommand{\CO}{\element[]CO}

\usepackage{printlen}
\usepackage{layouts}

\def\Htwo{H$_{\,2}$}

\newcommand{\kms}{$\,$km~s$^{-1}$}

\newcommand{\Jykms}{Jy\,km s$^{-1}$}
\newcommand{\mJyb}{mJy\,beam$^{-1}$}

\newcommand{\msun}{{${\rm M}_\odot$}}

\newcommand{\cmsq}{cm$^{-2}$}

\newcommand{\eg}{\mbox{e.g.}}
\newcommand{\ie}{\mbox{i.e.}}

\newcommand{\sauron}{{\texttt {SAURON}}}
\newcommand{\atlas}{{ATLAS$^{\rm 3D}$}} 

\newcommand{\ntr}{\mbox NGC~3100}
\newcommand{\nzr}{\mbox NGC~3095}
\newcommand{\dwn}{\mbox DwN}
\newcommand{\dww}{\mbox DwW}

\newcommand{\meer}{{MeerKAT}}

\newcommand{\vsys}{{$v_{\rm sys}$}\,}

\newcommand{\nhi}{{$N_{\rm H {\hskip 0.02cm \tt I}}$}}
\newcommand{\mhi}{{$M_{\rm H {\hskip 0.02cm \tt I}}$}}

\begin{document} 


   \title{The AGN fuelling/feedback cycle in nearby radio galaxies – \\ V. The cold atomic gas of NGC~3100 and its group}
   \titlerunning{\HI\ in NGC 3100}

        \author{F. M. Maccagni\inst{1,2}, I.    Ruffa\inst{3,4}, 
          A. Loni\inst{2,5}, I. Prandoni\inst{4}, R. Ragusa\inst{6,7}, D. Kleiner\inst{1,2}, P. Serra\inst{2}, E. Iodice\inst{6}, M. Spavone\inst{6}}

   \institute{ASTRON - Netherlands Institute for Radio Astronomy, Oude Hoogeveensedijk 4, 7991 PD, Dwingeloo, The Netherlands
        \and
        INAF -- Osservatorio Astronomico di Cagliari, via della Scienza 5, 09047, Selargius (CA), Italy
        \and
        Cardiff Hub for Astrophysics Research \&\ Technology, School of Physics \&\ Astronomy, Cardiff University, Queens Buildings, The Parade, Cardiff, CF24 3AA, UK
        \and
        INAF -- Istituto di Radioastronomia, via P.\ Gobetti 101, 40129 Bologna, Italy
        \and 
        Armagh Observatory and Planetarium, College Hill, Armagh BT61 9DG, UK
        \and
        INAF -- Osservatorio Astronomico di Capodimonte, via Moiariello 16, Napoli 80131, Italy
        \and Università di Napoli “Federico II”, Via Cinthia 21, Napoli 80126, Italy.
    \\
   \email{maccagni@astron.nl}
             }

   \date{Received May 01, 2023; accepted Month XX, XXXX}

 
  \abstract{We present Australia Compact Telescope Array (ATCA) 21-cm observations of the nearby low-excitation radio galaxy (LERG) {\mbox NGC~3100}. This is the brightest galaxy of a loose group and hosts a young ($\sim 2$ Myr) radio source. The ATCA observations reveal for the first time the presence of neutral hydrogen (\ion{H}{I}) gas in absorption in the centre of this radio galaxy, and in emission in two low-mass galaxies of the group and in a diffuse dark cloud in the proximity of {\mbox NGC~3100}. The sensitivity to low-column density gas ($N_{\rm H {\hskip 0.02cm \tt I}}\sim 10^{19}$~cm$^{-2}$) allows us to reveal asymmetries in the periphery of most the \ion{H}{I}-detected galaxies, suggesting that tidal interactions may be on-going. The diffuse cloud does not show a stellar counterpart down to  $27$ mag/arcsec$^2$ and could  be the remnant of these interactions. The analysis of the \ion{H}{I} absorption complex in {\mbox NGC~3100} indicates that the atomic phase of the hydrogen is distributed as its molecular phase (observed at arcsecond resolution through several carbon monoxide emission lines). We suggest that the interactions occurring within the group are causing turbulent cold gas clouds in the intra-group medium to be slowly accreted towards the centre of {\mbox NGC~3100}. This caused the recent formation of the cold circum-nuclear disk which is likely sustaining the young nuclear activity.}
	\titlerunning{\HI\ in \ntr}
	\authorrunning{F. M. Maccagni, et al.}
   \maketitle

\section{Introduction}
\label{sec:intro}
One of the most demanding requirements of modern galaxy formation theories is to reproduce the observed properties of early-type galaxies (ETGs) in the local Universe. How to prevent over-cooling of gas, why star formation (SF) quenches at late times, and which processes are responsible for the observed scaling relations between central super-massive black holes (SMBHs) and stellar bulges, are only some subjects of long-standing debates \citep[e.g.][]{Donofrio:2021}.

Active galactic nuclei (AGN) and their associated energetic output (i.e.\,feedback) are widely believed to play a crucial role in shaping galaxies over cosmic time \citep[e.g.][]{Harrison:2018}. One of the two main forms of AGN feedback is the so-called kinetic- (or jet-) mode feedback, where the bulk of the energy generated from the SMBH accretion process is channelled into collimated outflows of non-thermal plasma (i.e.\,the radio jets). The jets play an important role in heating the surrounding inter-stellar medium (ISM), possibly suppressing SF and maintaining the ETGs as passive, ``red and dead'' spheroids \citep[e.g.][]{Choi:2015}. However, the many details of how radio jets are powered and interact with their surroundings are still not fully understood~\citep[e.g.][]{Harrison:2018,Hardcastle:2018}.

Important insights into this matter can come from systematic investigations of a particular class of AGN, known as low excitation radio galaxies (LERGs). LERGs are - by number - the dominant radio galaxy population in the local Universe \citep[e.g.][]{Hardcastle:2007}, predominantly hosted by red, massive ($M_{*} \geq 10^{10}~M_{\odot}$) ETGs \citep[e.g.][]{Best:2012,Miraghaei:2017}. These AGN eject most of the energy through collimated radio jets, inducing kinetic feedback into the ISM. 

Detailed spatially-resolved studies of nearby LERGs are key to probe the formation and evolution of massive ETGs in relation to the fuelling/feedback cycles associated with their nuclear activity.

ETGs are typically devoid of cold ($T\lesssim 10^2 K$) gas in their centres. However, over the past decade, significant amounts of cold molecular gas and dust sufficient to sustain low accretion rates of LERGS have been found in the central regions of ETGs (e.g.\,M$_{\rm H_{\rm 2}} \sim 10^7 - 10^{10}$~M$_{\odot}$; \citealp[e.g.][]{Ocana:2010,Prandoni:2010,Ruffa:2019b,Tamhane:2022}). The origin of the observed cold material is still hotly debated: it may be either internally generated (for example, through stellar mass loss or hot halo cooling) or externally accreted (through for example, environmental effects such as mergers and tidal interactions).

In clusters and groups both galaxy encounters, tidal interactions as well as hydrodynamical effects ram pressure stripping can be responsible for directly driving the gas from the inter-galactic medium into the centre of galaxies, triggering the nuclear activity~\citep[e.g.][]{Almeida:2012,Poggianti:2017}. Moreover, these same phenomena inject turbulence in the environment, which may also generate cooling of gas in the hot halo of galaxies and consequent infall onto the central SMBH, either directly and smoothly \citep[e.g.][]{Negri:2014} or - more realistically - after chaotic cooling (as predicted in chaotic cold accretion models, CCA; \citealp[e.g.][]{Gaspari:2015,Gaspari:2017}). Filamentary or blob-like cold gas structures, likely reminiscent of condensing multi-phase gaseous clouds, have been frequently observed extending from the hot halo onto the centre of nearby LERGs in high-density environments, supporting the scenario described above~\citep[e.g.][]{Tremblay:2018,Nagai:2019,Maccagni:2021,Tamhane:2022}.

In high luminosity radio sources ($L_{\rm 1.4 GHz} \gtrsim 10^{24}$ W Hz$^{-1}$), a connection between the evolution and activity of the SMBH, the fuelling cold gas, and the surrounding dense environments is suggested by the tight correlations observed between their hot X-ray emitting halos, radio jet powers and molecular gas masses~\citep[\eg ][]{Ineson:2015,Pulido:2018,Babyk:2019}.

The incidence of environmental phenomena in less powerful radio galaxies ($L_{\rm 1.4 GHz}\gtrsim 10^{22-23}$ W Hz$^{-1}$) is still highly debated~\citep[\eg ][]{Ching:2017}. Simulations suggest that these sources are hosted by less massive dark matter haloes~\citep[$\sim10^{12}$\msun][]{Thomas:2021}, where the environment of galaxies is typically less extreme than in clusters. In these galaxies, rotation is often not negligible and it is indeed still unclear what is the role of turbulence in funnelling the gas onto the SMBH, compared to secular processes occurring within the galaxy and minor interactions in the halos \citep[e.g.][]{Sabater:2015,Gaspari:2015,Davis:2019,Temi:2022}. Key answers in this regard can come from the study of the distribution and kinematics of the cold gas in LERGs on all scales, from the circum-nuclear to the circum-galactic environments. 

Radio interferometers can carry out these studies in nearby AGN by observing the neutral atomic hydrogen (\HI) gas at 21-cm wavelengths. This is often found in the centre of active galaxies~\citep[\eg ][]{Morganti:2001,Maccagni:2017,Curran:2018}, traces the emission of cold gas in ETGs beyond the central regions in galaxies and is the most abundant gas phase in the inter-galactic medium, allowing us to directly probe the presence of interactions between nearby ETGs \citep[as, for example, shown by the \sauron\ and \atlas\ surveys][]{Oosterloo:2010,Serra:2012}.

Diffuse (\nhi$\sim 10^{19}$\cmsq) \HI\ clouds have been associated with outflowing gas in several nearby radio AGN, through the detection of high-velocity absorption components against the radio jets~\citep[see ][for a census]{Morganti:2018}. However, the sensitivity of past radio interferometers limited these studies to absorption features, which depend strongly on the distribution of the underlying radio continuum, and are therefore very limited in determining the full extent of the detected gas distribution.

The Australia Telescope Compact Array (ATCA) has a wide field of view ($\sim1$ deg$^2$), which is crucial to study the \HI\ gas from the circum-nuclear to the circum-galactic regions of nearby AGN, and also a combination of short baselines which make it sensitive to the low-column density neutral hydrogen (even though at low resolutions, $\sim 30\arcsec - 90$\arcsec). Several ATCA \HI~observations made it possible to obtain promising results in the study of the cold gas supply mechanism in nearby AGN. For example, in the LERG \mbox{NGC~612} ~\citet{Emonts:2008} discovered an enormous (140-kpc wide) \HI\ disk connected by a prominent low-column density ($\lesssim 8\times10^{19}$~\cmsq) bridge with its \HI-rich companion galaxy \mbox{NGC~619}, indicating that a past interaction between the two galaxies may have channeled large ($10^9$~\msun) amounts of cold gas into this LERG. Deep ATCA \HI\ observations have also showed that the large-scale \HI\ disk of the radio galaxy \mbox{PKS 1718-649} was likely formed through a gas rich merger, but that this event is disconnected to the more recent funnelling of cold gas into the centre and consequent triggering of the AGN, as suggested by the \HI\ detected in absorption~\citep[][]{Maccagni:2014}.

The complex nature of SMBH feeding mechanisms and the wealth of observational evidence that can be derived from a single object clearly indicate the need to expand these studies to larger samples of LERGs. Atacama Large Millimeter/submillimeter Array (ALMA) CO(2-1) observations~\citep[e.g.][]{Ruffa:2019a} have shown that (sub-)kpc molecular gas disks are very common in LERGs. We are carrying out the first systematic, spatially resolved, multi-component (stars, hot/warm/cold gas, dust and radio jet) study of a small but complete sample of 11 LERGs in the southern sky. A full description of the sample can be found in \citet[][]{Ruffa:2019b}. Although the bulk of this gas is found to be in ordered rotation, low-amplitude perturbations and/or non-circular motions are ubiquitous \citep[][]{Ruffa:2019a}, which likely drive the gas towards the SMBH.
The origin of the circum-nuclear disks in these sources is still very unclear. Most of them are located in poor environments (i.e.\,poor groups, pairs or even isolated). For this reason, we exploit the wide field of view and high sensitivity to diffuse \HI\ of the ATCA to measure the total mass and distribution of the cold gas that may live in these galaxies and their environment, and conclusively determine which mechanisms (\ie\ external such as interactions, or internal such as CCA) bring the gas from the circum-galactic to the circum-nuclear regions, thus sustaining their nuclear activity.  

We obtained deep ATCA \HI\ observations of all sources in 2020. In this paper, we present the wide-field and high-sensitivity \HI\ observations of the nearby ($38$ Mpc) source, \ntr~\footnote{Throughout this work we assume a standard $\Lambda$CDM cosmology with H$_{\rm 0}=70$\,km\,s$^{-1}$\,Mpc$^{\rm -1}$, $\Omega_{\rm \Lambda}=0.7$ and $\Omega_{\rm M}=0.3$. This gives a scale of $182$~pc/$''$, at the redshift of NGC\,3100 ($z=0.008$; \citealp[][]{Ruffa:2019b}).}. Coupling \HI\ data with the deep optical image of \ntr\ taken in the context of the VST Early-type GAlaxy Survey \citep[VEGAS; \eg][]{capaccioli:2015,iodice:2021} and the previous ALMA CO observations \citep[][]{Ruffa:2019b,Ruffa:2022}, we identify and characterize the processes that brought the cold gas from the circum-galactic environment into the circum-nuclear regions of this LERG and likely fuel its nuclear activity.

The paper is structured as follows. In the following section we provide a brief description of the main properties of \ntr. In Section~\ref{sec:obsDR} we describe the ATCA observations and data reduction. The data analysis is presented in Sections~\ref{sec:hiRes} and~\ref{sec:hiAbs}. We discuss the results in Section~\ref{sec:disc}, before summarising and concluding in Section~\ref{sec:conclusions}.

\section{The target: \ntr}
\label{sec:ntr}
\begin{figure*}  
	\centering
	\includegraphics[width=\textwidth]{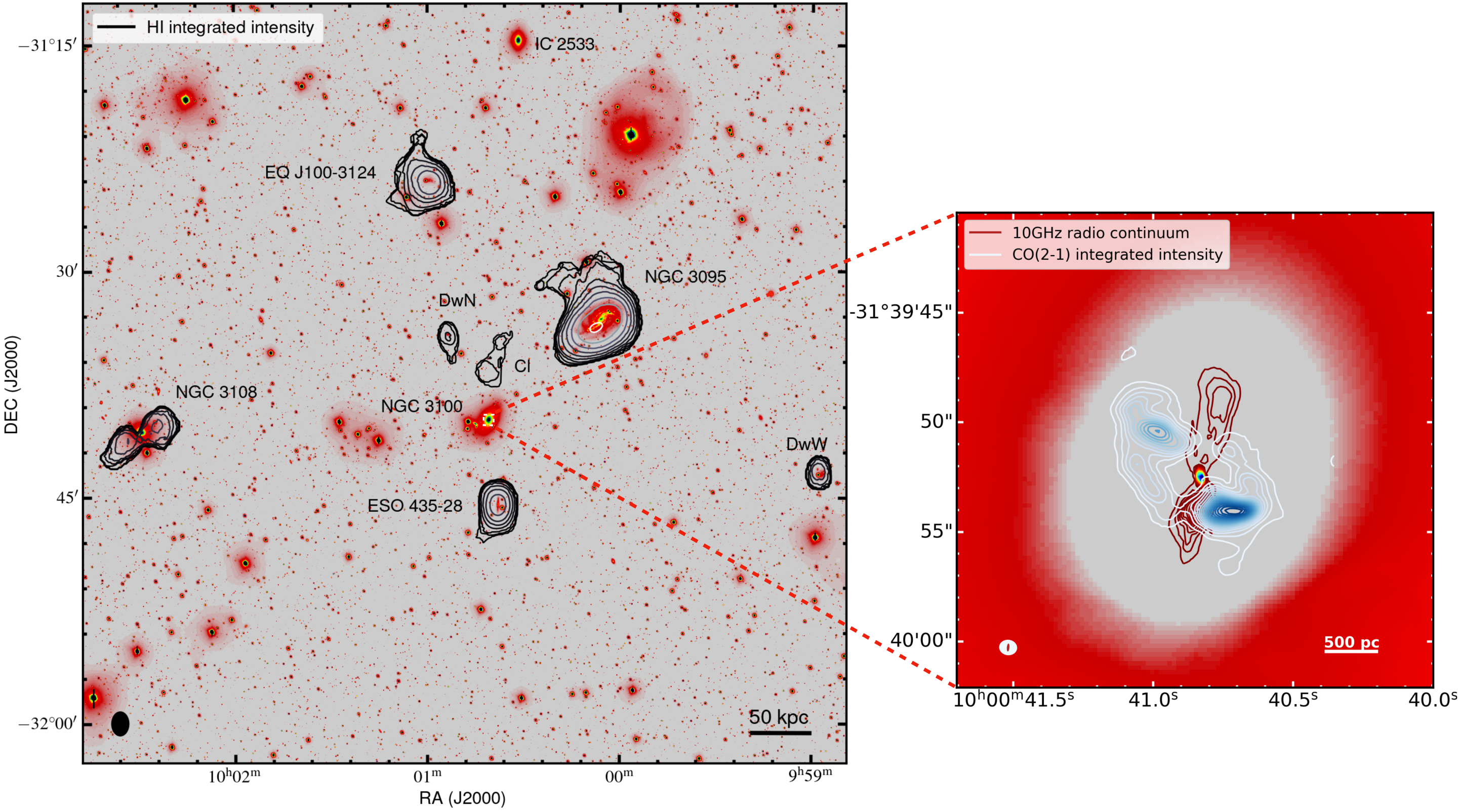}	
\caption{{\em Left panel} Primary beam-corrected \HI\ contours overlaid on a r-band image from VEGAS of the NGC3100 group. The contour levels are $5\sigma \times 2^n$\Jykms. The lowest contour corresponds to $1.1\times 10^{19}$\cmsq. {\em Right panel}: VEGAS r-band map zoomed in the inner $15''\times15''$ ($\approx2.5\times2.5$~kpc$^{2}$). The colourscale is inverted compared to the left panel. ALMA CO(2-1) integrated intensity contours from \citet{Ruffa:2019a}, and radio continuum contours from JVLA data at 10~GHz \citep{Ruffa:2020} are overlaid in white-to-blue and red contours, respectively. Contours are drawn at 1,3,9... times the 3$\sigma$ r.m.s. noise level. The \HI\ (left panel), CO and radio continuum (right panel) Gaussian restoring beams are shown in the bottom-left corners. A scale bar is shown in the bottom-right corner of each panel.}
	\label{fig:vegasHI}%
\end{figure*}

\ntr\ is a late S0 galaxy, characterised by a patchy dust distribution and a bright nuclear component \citep[][]{Laurikainen:2006}. It is the brightest galaxy of a loose group and forms a pair with the barred spiral galaxy \mbox{NGC\,3095}, located at a projected linear distance of $\approx 95$~kpc (Fig.~\ref{fig:vegasHI}). Multi-frequency radio continuum data show that \ntr\ hosts the core-double lobe radio source PKS~0958-314 \citep[$P_{1.4 \rm GHz} = 10^{23}$ W Hz$^{-1}$, e.g.][]{Ekers:1989,Ruffa:2019b}, with jet power in the range $10^{43}-10^{44}$~erg~s$^{-1}$ \citep[][]{Ruffa:2022}. The radio source has a total linear extent of $\approx 1.8$\,kpc and there are no indications of radio emission on larger scales \citep{Ruffa:2020}. According to the jet size-age correlations \citep[e.g.][]{ODea:2021}, the AGN can be considered in an early phase of its evolution (less than 2~Myr).
\ntr\ was observed with ALMA during Cycles 3 and 6 and resolved with arcsecond resolution in three different CO rotational transitions (up to $J=3$). The carbon monoxide is distributed in an edge-on ($i=60^\circ$, $PA=220^\circ$) rotating ($v_{\rm rot} = 345$~\kms) disk extending for $\approx2$ kpc in the centre of the galaxy. The radio jets expand in a direction close to the plane of the disk. A detailed view of the CO and radio emission in \ntr\ is given in the zoom-in panel of Fig.~\ref{fig:vegasHI}. The 3D kinematic modelling of the molecular gas disk shows that overall the disk is regularly rotating, but several clouds have non-circular motions \citep{Ruffa:2022}. These clouds likely form both inflow streaming motions and a jet-induced outflow in the plane of the CO disk, with $v_{\rm max}\approx 200$~km~s$^{-1}$ and $\dot{M}\lesssim 0.12$~M$_{\odot}$~yr$^{-1}$ \citep[][]{Ruffa:2022}. In this paper, we show for the first time that \HI\ gas is present in the central regions of \ntr\ and its surrounding large-scale environment, and how the study of its distribution and kinematics allows us to determine how the cold gas is supplied to the AGN.

\section{Observations and data reduction}
\label{sec:obsDR}

\subsection{1.4 GHz ATCA observations}
\label{sec:atcaObs}
The ATCA  observations of \ntr\ were taken in three separate runs of 12 hours each, with different array configurations (see Table~\ref{tab:atcaObs} for details). The observations were centred at the \HI\ frequency of the systemic velocity of the galaxy, $1409.13$~MHz. The Compact Array Broadband Backend (CABB) of the telescope provides a total bandwidth of 64 MHz over 2048 channels (dual polarisation). At the \HI\ frequency, in the Local Universe, the channel width is $\approx 6.6$\kms.

The data were calibrated and cleaned using {\tt MIRIAD}~\citep[][]{Sault:1995,Sault:2011}. A continuum image was produced using the line-free channels. At the resolution of our observations ($\approx 8$\arcsec), the continuum is unresolved with a flux density of $S_{\rm 1.4 \,\, GHz} = 491.5$ mJy. The continuum was subtracted by fitting the visibilities in the line free channels with a first order polynomial. 

We detected \HI\ in emission in the field of view ($\sim 1$ deg$^2$) of the observations and in absorption in the centre, against \ntr. We produced two different datacubes to best retrieve the information from both the \HI\ detected in emission and in absorption. Table~\ref{tab:atcaObs} reports a complete list of parameters of the two datacubes. For the emission, to increase the signal-to-noise ratio, the final datacube considers visibilities only from five antennas, thus excluding the longest baselines. The restoring beam has a size of 66\arcsec$\times$41\arcsec. To increase the column density sensitivity we smoothed the cube to a final PSF of 99\arcsec$\times$86\arcsec and binned three spectral channels together. The noise of the datacube is $0.89$~\mJyb at $19.9$\kms\ channel width. Following the equations given in \citep[][]{Meyer:2017}, at the achieved sensitivity, the minimum detectable column density ($3\sigma$) is $6.9 \times 10^{18}$~\cmsq\ and the minimum detectable mass is $6\times 10^6$~\msun. 
We used {\tt SoFiA}~\citep[][]{Serra:2015} to determine the reliable \HI\ emission of each source. By smoothing and clipping, we convolved the input cube with a set of kernels and detected emission above 3.5 times the local rms noise level of each kernel. We created the \HI\ density map and velocity field using the same software suite. We confirmed the characterisation of the detections also by visually inspecting the datacube. 

The optical depth and, consequently, the column density sensitivity of the \HI\ detected in absorption depends on the noise of the observations, but also on the flux of the background continuum source. Hence, in absorption it is possible to reach similar column density sensitivities at higher spatial resolutions than in emission. Keeping all six antennas of the ATCA, we generated a datacube with a restoring beam of 21\arcsec$\times$19\arcsec\ and binned three channels together. The rms noise level in the absorption datacube is $0.55$~\mJyb. Against the $1.4$~GHz continuum emission of \ntr, this corresponds to a $3\sigma$ column density detection limit of $1.2 \times 10^{19}$~\cmsq, assuming a spin temperature of the gas $T_{\rm spin}=100$~K and that the \HI\ fully covers the background continuum spanning over three velocity channels ($19.7$~\kms). Taking the background radio-continuum function as the maximum extent of the \HI\ detected in absorption, the minimum detectable mass is $3.4 \times10^5$~\msun.

\begin{table}[tbh]
        \caption{ATCA observations and data products}
        \centering
        \label{tab:atcaObs}
        \begin{tabularx}{\columnwidth}{l X }  
                \hline\hline                                                         
                Date & 12-12-20,\,15-01,\,10-02-21\\
                	Proj. ID & C3364\\
                	Telescope configurations & 750C,1.5A, EW352 \\
                	Pointing Centre (J2000) & $10^{\rm h}\,00^{\rm m}\,40.8^{\rm s}$, $-31^{\rm d}\,39^{\rm m}\,52{\rm s}$\\
                	Bandwidth       & 64 MHz\\
                	Channel width   &  6.6\kms \\
                	Number of antennas & 5 (6) \\
                	Time on target & $12+ 12 + 12$ hrs \\
                	HI cube weighting & 0.5 \\
                	HI cube restoring PSF & 66\arcsec$\times$41\arcsec (21\arcsec$\times$19\arcsec)\\
                	HI cube final PSF (*) & 100\arcsec$\times$86\arcsec  (21\arcsec$\times$19\arcsec)\\
                	HI cube r.m.s. noise & 0.89 (0.55)~\mJyb \\
                	3$\sigma$ column density (3-ch) &  $6.9\times10^{18}$  ($1.25\times10^{19}$)~\cmsq \\
                \hline                           
        \end{tabularx}
         \tablefoot{Numbers within parenthesis refer to the datacube used for the analysis of the \HI\ absorption.
         (*) To increase the column density sensitivity we smoothed the cube to a final PSF, and binned 3 channels toghether, hence the r.m.s. noise is given over channels 19.9~\kms\ wide, see Sect.~\ref{sec:atcaObs}.}
\end{table}

\subsection{Deep optical images of the \ntr\ group}
New and deep images, in the optical $g$ and $r$ bands, for the \ntr\ group have been obtained with OmegaCam at VLT Survey Telescope (VST), as part of the VST Early-type GAlaxy Survey~\citep[VEGAS, see][]{iodice:2021} between December 2019 and January 2020. Observations cover a large area ($\sim 1 $~square degree) around the group, with a pixel scale of 0.21 arcsec/pixel \citep{kuijken2011Msngr.146....8K}.

The images are reduced by using the dedicated {\it AstroWISE}  pipeline \citep[see][for details.]{McFarland2013,Venhola_2017,Venhola2018}. Data were acquired in dark time during the ESO run P104 (run ID: 0104.A-0072[B]). These data have been analysed to perform the surface photometry of all group members and constrain the total amount of intra-cluster light in this environment, as published by \citet{Ragusa2023}.

Thanks to the long integration time ($\sim3.5$~hours and $\sim2.75$~hours on source for the g and r filters respectively), with these new imaging data we are able to map the light distribution down to the faint surface brightness levels of $\mu_g \sim 28.9$~mag/arcsec$^2$, 
$\mu_r \sim 27.3$~mag/arcsec$^2$, in the $g$ and $r$ bands, respectively. 
Moreover, the large field of view of OmegaCam allows us to cover the whole area observed with ATCA ($\sim 1.5$ deg$^2$). The VEGAS image of the \ntr\ group, in the $r$ band is shown in Fig. \ref{fig:vegasHI},  with overlaid the contours of the \HI\ detected in emission and in absorption against the radio emission of \ntr.  As discussed in Sec.~\ref{sec:group}, combining deep optical images with the \HI\ density is fundamental to detect any, and faint, star-light counterpart of the gas and, therefore, to address the possible origin for the latter component. 

\section{Distribution and kinematics of the cold gas}
\label{sec:hiRes}
The ATCA \HI\ observations of \ntr\ provide an unprecedented view of its gas content and environment. For the first time, \HI\ is revealed in different galaxies of the group of \ntr. The main panel of Fig.~\ref{fig:vegasHI} shows the distribution of the \HI\ overlaid on the deep optical image of the \ntr\ group made by VEGAS. The group of \ntr\ has $1.18\times10^{10}$\msun\ of \HI\ distributed among six satellites within $1.5$ deg$^2$ from the bright radio galaxy.  In particular, we detect \HI\ gas associated to the disk of NGC 3095, NGC 3108, {\mbox ESO 435-28}, {\mbox EQ J100-3124}, and - for the very first time - in  two dwarf galaxies (\mbox{DwN}, {\mbox DwW}), and in a cloud located in the sky area between \ntr\ and its companion, \mbox{NGC 3095} ({\em Cl}). Most of the detected \HI\ disks show morphological asymmetries. Moreover, we discover \HI\ absorption against the radio source in \ntr. The main properties of the \HI\ detections are summarised in Table~\ref{tab:masses}.  

Previous \HI\ observations already detected gas in some of the sources of the \ntr\ group, but never associated to the AGN. In particular, \nzr\ is detected in the \HI\ Parkes All Sky Survey~\citep[HIPASS, ][]{Meyer:2004,Zwaan:2004}. \dwn\ and the cloud {\em Cl} are both within the HIPASS PSF (15.5\arcmin) and thus formally indistinguishable from \nzr. {\mbox ESO 435-28} was already known to have some \HI, detected by single dish observations~\citep[][]{Theureau:1998}. ATCA observations a factor 1.4 less sensitive than the one presented in this paper, at comparable resolutions, previously observed the \HI\ in \mbox{NGC 3108}~\citep[\eg][]{Oosterloo:2002,hau:2008,serra:2008}. The same observations also discovered \HI\ in {\mbox EQ J100-3124}, referring to this source as `anonymous galaxy' \mbox{A1000--31}. Our results confirm the \HI\ mass and systemic velocity of the latter, while we recover only partially the \HI\ mass of \mbox{NGC 3108} reported by \citet{Oosterloo:2002} ($4.6\times 10^9$~\msun, at the distance of 53 Mpc). This is likely due to the fact that this galaxy is $\sim 30$\arcmin\ away from \ntr, at the edge of the field of view of our ATCA observations, where the sensitivity of the primary beam drops below $10\%$. Likely, also the \HI\ detection in DwW is affected by this issue. 

The \HI\ in \dwn\ and \dww\ has low-column densities ($\sim 5\times 10^{19}$\cmsq) and \dwn\ is the galaxy with the lowest \HI\ mass detected in emission, \mhi~$=8\times 10^7$~\msun. The most interesting novel discovery is the extended ($4$\arcmin) \HI\ cloud ({\em Cl}) 38 kpc away from \ntr, between \dwn\ and \nzr. Its \HI\ is very diffuse spanning column densities between $6 \times 10^{18}$~\cmsq\ and $3 \times 10^{19}$~\cmsq. The cloud is detected over three spatial beams and seven spectral channels, with an average $S/N \gtrsim 3.5$ in each channel. The VEGAS observations suggest that no stellar counterpart is associated with this cloud (further details are given in Sect.~\ref{sec:group}).

\begin{figure}
	\centering
	\includegraphics[width=\columnwidth]{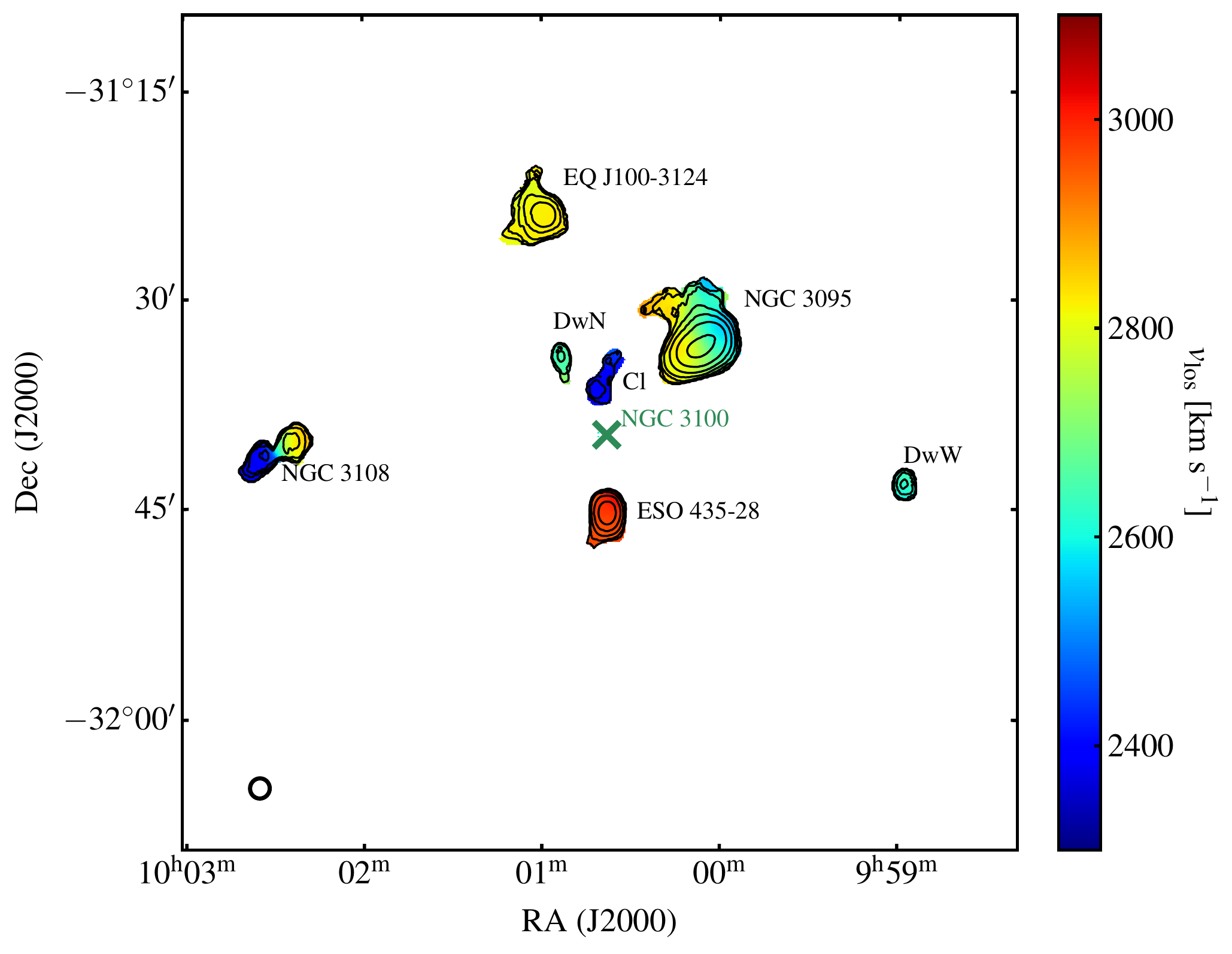}	
	\caption{Velocity field of the \HI\ detected in the group of \ntr\ (whose location is marked by a green cross, \vsys$=2641$\kms) Contour levels are as in Fig.~\ref{fig:vegasHI}.}
	\label{fig:mom1}%
\end{figure} 

\begin{table*}[tbh]
        \caption{\HI\ properties of the sources detected in the \ntr\ group.}
        \centering
        \label{tab:masses}
        \begin{tabularx}{\textwidth}{X c c c c c c} 
                \hline\hline                                                         
                Source Name	& Coordinates &\vsys (\HI)   & $w50$  & $w80$   & F$_{\rm HI}$   &\mhi   \\
                	& [J2000] & [\kms]  &  [\kms] & [\kms]  & [\Jykms]   & [\msun]   \\
                \hline
                \ntr$(\dag)$		&10:00:40.80,-31:39:52.3&	2641   & 	178  &	397	&	-0.056 (*)  &$1.1\times 10^6$\\
                            		&                       &	       & 	      &		&	            &$5.1\times 10^7$(**)\\
                \mbox{ESO~435-028} 	&10:00:37.93,-31:45:29.1	& 2979 	& 128 & 158 & 2.5 & $8.6\times10^8$		\\
                \mbox{NGC~3095}		&10:00:05.74,-31:33:22.0	& 2706 	& 338 & 365 & 22  & $7.5\times10^9$		\\
                \mbox{EQJ100-3124}	&10:01:00.42,-31:24:01.2	& 2820 	& 67.1& 101 & 4.1 & $1.3\times10^9$		\\
                \mbox{NGC~3108}		&10:02:27.52,-31:42:41.3	& 2626 	& 535 & 547 & 3.4 & $1.1\times10^9$		\\     
                \mbox{DwN}			&10:00:53.54,-31:34:13.1	& 2655 	& 86.0& 114 & 0.25& $8.0\times10^7$		\\     
                \mbox{DwW}			&09:58:58.19,-31:43:28.3	& 2979 	& 128 & 158 & 0.84& $8.6\times10^8$		\\     
                \mbox{Cl}			&10:00:40.58,-31:36:41.7	& 2401 	& 41.0& 59.2& 0.29  & $1.0\times10^8$		\\     
                \hline                           
        \end{tabularx}
        \tablefoot{$(\dag)$ detected in absorption.\\
        (*) corresponds to an optical depth of $\tau=0.114$\\
        (**) mass range estimated with different assumptions on the real extent of the \HI\ gas detected in absorption, see Sect.~\ref{sec:hiAbs} for further details.}
\end{table*}

The \HI\ systemic velocities of the gas rich galaxies (shown in Table~\ref{tab:masses}) are all within $300$~\kms\ from \ntr, confirming that they all likely belong to the same group.
The velocity field in Fig.~\ref{fig:mom1} shows that \mbox{NGC 3095} and DwN have velocities very similar to \ntr\ (within $60$\kms), while the diffuse cloud is blue-shifted by $200$\kms.  In Section~\ref{sec:group} we combine this information with the deep photometry to understand the formation history of this group and relate it to the gas content of its members.

\section{The \HI\ in \ntr\ detected in absorption}
\label{sec:hiAbs}

\begin{figure}[tbh]
	\centering
	\includegraphics[width=\columnwidth]{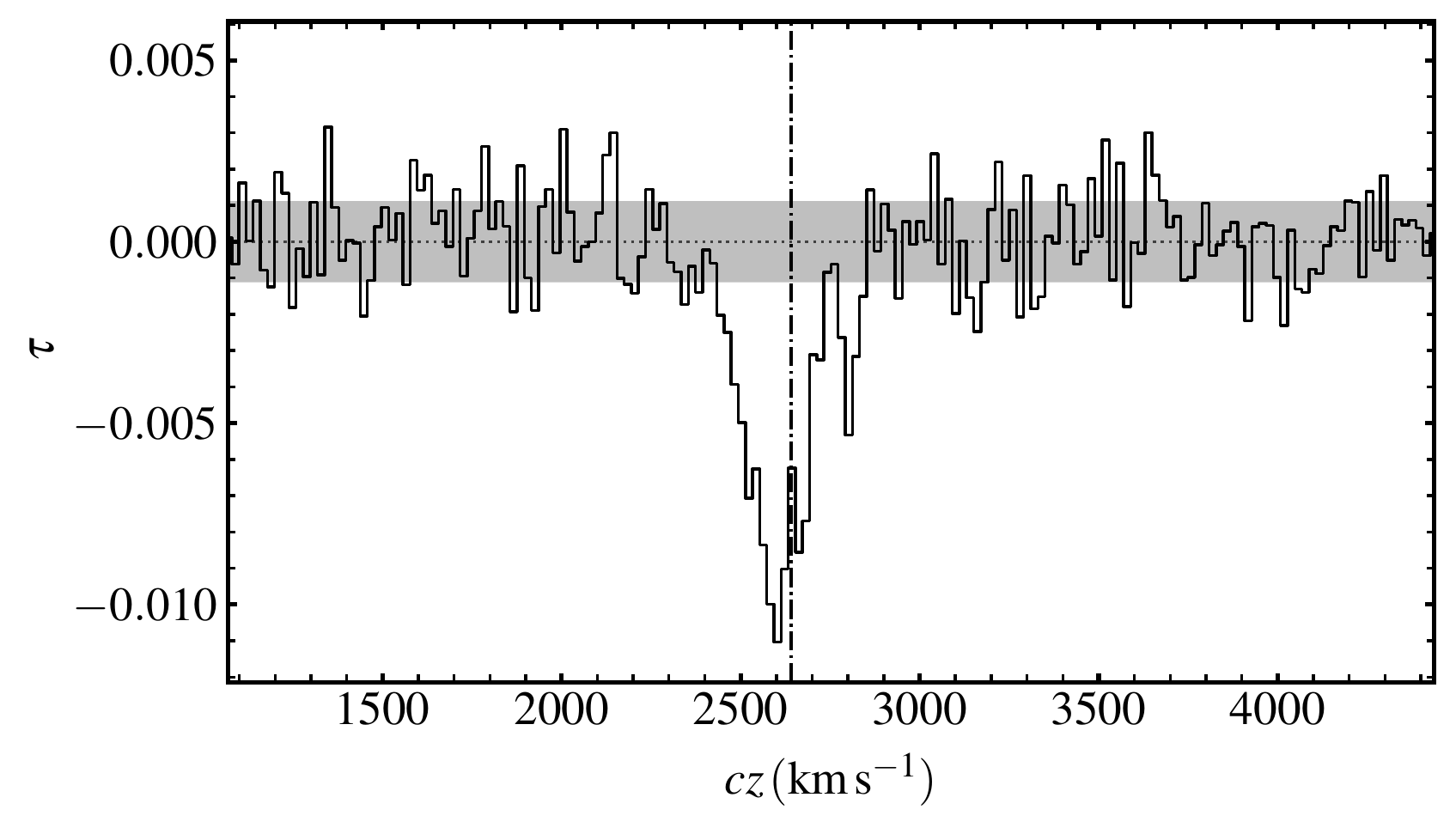}	
	\caption{\HI\ absorption detected in the centre of \ntr\ (with a PSF of 21\arcsec$\times$19\arcsec) against the unresolved radio continuum emission of the AGN ($S_{\rm 1.4 GHz}\sim 490$ mJy). The dashed-dotted vertical line marks the systemic velocity of the source (\vsys$=2641$~\kms), the gray shaded area shows the average noise across in the datacube.}
	\label{fig:absorption}%
\end{figure}

Figure~\ref{fig:absorption} shows the \HI\ absorption profile of \ntr. The complex has a main broad line peaking close to the systemic velocity ($\tau = 0.0119$) and a second narrower (full width at zero intensity FWZI$ = 73$\kms) redshifted component ($v_{\rm peak}-v_{\rm sys}\sim+200$\kms). The peak of the line is close to the systemic velocity ($v_{\rm peak}-v_{\rm sys}\sim-31$\kms), which is  indicative of an absorption line tracing a circum-nuclear disk~\citep[\eg][]{Gereb:2015,Maccagni:2017}. The full-width at half maximum ($w50=397$~\kms) is within the range of rotational velocities of \ntr, as traced by its circumnuclear disk of molecular gas~\citep{Ruffa:2019b}. Nevertheless, the FWZI of the overall absorption complex is $570$\kms, which is greater than the FWZI of the molecular gas detected in the centre of this AGN~\citep[$\approx 430$\kms; see, for example, Fig. 5 of][]{Ruffa:2022}. The FWZI of the CO is similar to the full width of the broad component of the \HI\ absorption. This suggests that only part but not all of the \HI\ absorption is tracing the same neutral atomic phase of the circum-nuclear disk seen in the CO. Further information on the distribution of the \HI\ traced in absorption and its relation with the molecular gas is given in Sect.\ref{sec:hico}. 

In our observations, the radio jets and core of \ntr\ are unresolved, Assuming that the \HI\ completely covers the background continuum emission (covering factor $c_f=1$), and that their spin temperature is $100$~K, the integrated column density of the absorption complex is \nhi\ $=4.43\times10^{20}$\cmsq. The total mass of the \HI\ traced by the absorption may be estimated as \mhi~$\pi r^2\cdot$\nhi, making some assumption on the radius of the absorbing clouds ($r$). An upper limit on the mass is obtained assuming that the \HI\ is covering the entire PSF of the observations ($21$\arcsec$\times 19$\arcsec, $\approx 3.8\times 3.8$ kpc) and is \mhi=$5.13\times 10^{7}$\msun. A lower limit on the mass is derived assuming that the absorption cannot occur beyond the extent of the radio jets as given by their emission at $10$ GHz~\citep[$10'',\sim 1.8$~kpc,][]{Ruffa:2020}. This implies a total mass of the absorption complex of \mhi~$=1.1\times 10^{6}$~\msun. 

\section{Discussion}
\label{sec:disc}

\subsection{The \HI\ in the \ntr\ group}
\label{sec:group}

The novel \HI\ detections in combination with the deep VEGAS optical observations allow us to identify two new \HI\ rich dwarf galaxies in the group of \ntr: DwN, DwW, and to detect several asymmetries in the outskirts of the \HI\ disks of all other sources close to \ntr\ (\ie\ \nzr, ESO~425-28, EQ~J00-3124). Here, we investigate the \HI\ and stellar properties of these sources in relation to the assembly history of the group. 
The overlay between the combined {\em g, r} color image from the VEGAS observations and the \HI\ distribution (Fig.~\ref{fig:vegasHI} and~\ref{fig:cloudOptical}) shows that while the detected \HI\ disks have asymmetries in their outskirts, whereas their stellar bodies are not majorly disturbed. This suggests that the group galaxies did not encounter a recent major merger, but that weak tidal interactions are likely ongoing.

We also detect for the first time diffuse \HI\ emission at the outskirts of \ntr\ ($38$ kpc away from the centre).  The deep optical imaging shows that no stellar counterpart is present at the coordinates of the \HI\ cloud, or in its proximity, down to the surface brightness magnitude limit of $27.3$ mag~arcsec$^{-2}$, see Fig.~\ref{fig:cloudOptical}). Hence, a dark \HI\ cloud of $\sim 1\times 10^8$\msun\ seems to be located approximately half-way between \ntr\ and its companion \nzr. The cloud has an asymmetrical shape and it could be the remnant of a past interaction between these two galaxies. Nevertheless, even though {\em Cl} seems to point towards the northern tail of \nzr, it is blue-shifted with respect to it and to \ntr\ of $150$ and $300$~\kms, respectively, and no further \HI\ emission is found between the galaxies (see also Fig~\ref{fig:mom1}, and Table~\ref{tab:masses}).

\begin{figure}[tbh]
	\centering
	\includegraphics[width=\columnwidth]{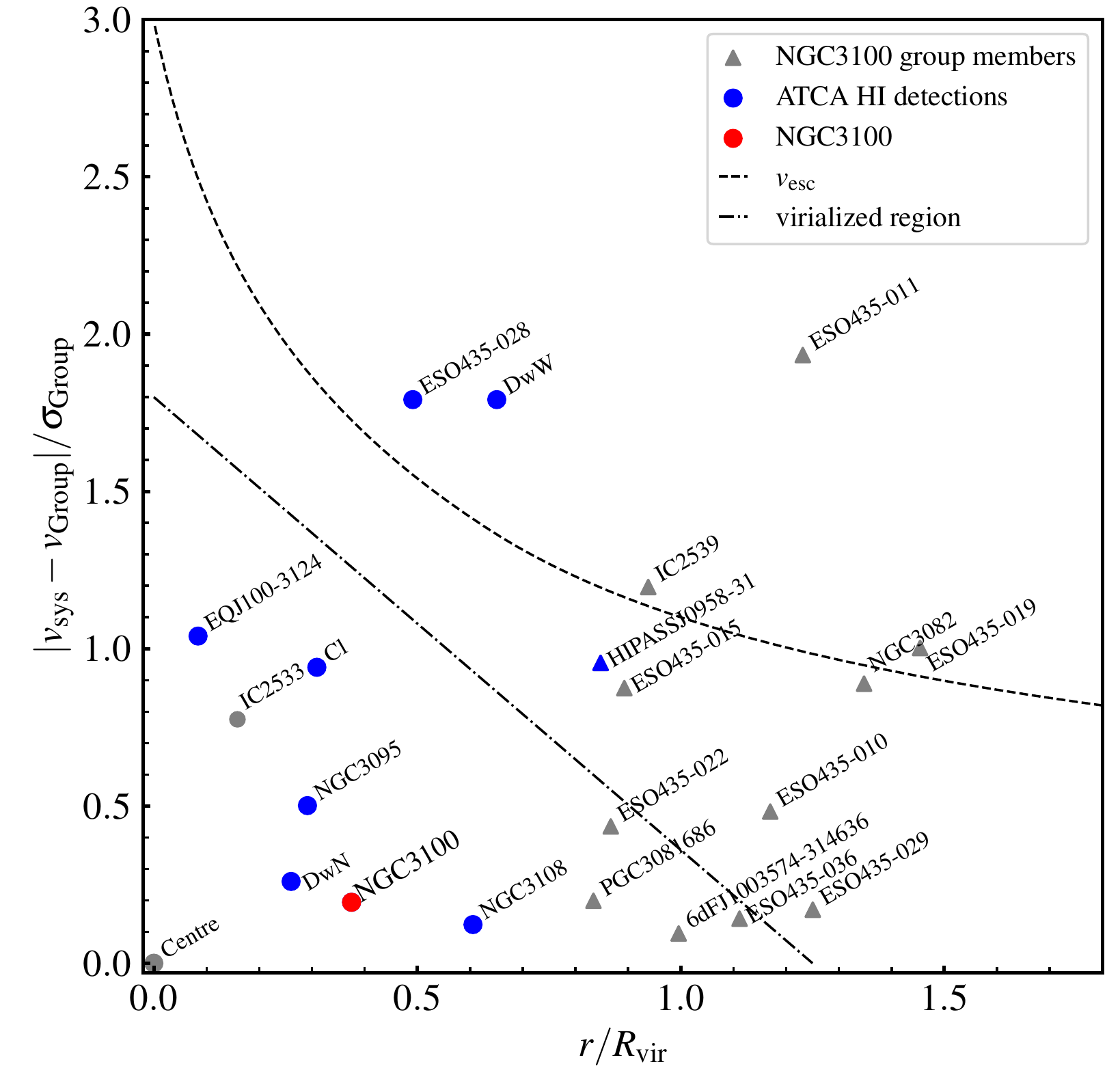}	
	\caption{Phase-space diagram of the \ntr\ group. \ntr\ is shown in red, while \HI\ detections are shown in blue, non-detections in grey. Triangles indicate group members from the ~\citet[][]{kourkchi:2017} catalog, located outside the field of view of ATCA. The dotted line shows the caustic curve defined by the escape velocity of the group. Interestingly, most \HI\ detections are located within the virialised region of the group (dashed-dotted line). The proximity of \mbox{NGC 3095} and DwN suggests a recent interaction with \ntr.}
	\label{fig:phaseSpace}%
\end{figure} 

Further information on the origin of such \HI\ cloud and on the possible assembly history of the \ntr\ group can be obtained from their location in the phase-space diagram~\citep[\eg][]{Jaffe:2015,Rhee:2017}. This diagram relates the projected distance from the centre of the group (normalized by the virial radius of the group) with the systemic velocity of each galaxy with respect to the group centre (normalized by its velocity dispersion). In Fig.~\ref{fig:phaseSpace} we show the projected phase-space diagram of the \ntr\ group. In addition to the \HI\ detections, we show the group members identified in the `galaxy groups within 3500~\kms ' catalog by~\citet{kourkchi:2017}, together with the virialized area of the group and the caustic curves defined by its escape velocity (which depends from its virial mass, radius and velocity dispersion, $M_{\rm vir} = 6.6\times10^{13}$~\msun, $R_{\rm vir} = 0.53$ Mpc, $\sigma_{\rm group}=211$~\kms, respectively). We highlight that in the ~\citet{kourkchi:2017} catalog the centre of the group is not on the \ntr\ but is shifted towards the north (Ra =10:00:40.800, Dec=-31:39:52.30, $v_{\rm rad}=2600 $\kms). Moreover, the ATCA observations detect \HI\ in all group members visible in the field of view (marked by circles in the Figure) except than in \mbox{IC 2533} in the North.

Different simulations~\citep[][]{Jaffe:2015,Smith:2015,Rhee:2017} show that in the virialized region it is more likely to find old members of the group between which interactions have already occurred, compared to recent infallers which are typically located between this region and the caustic lines. This has been observed in denser environments than the \ntr\ group, such as the Fornax cluster~\citep[][]{Loni:2021} and Virgo~\citep[][]{Jaffe:2016}. Interestingly, in the loose group of \ntr\ most \HI\ detections are found in the virialized region, while in clusters and dense groups the \HI\ is typically located in the region of the recent infallers~\citep[][]{Jaffe:2015,Loni:2021}. The diagram also suggests that not only \nzr\ may have interacted with \ntr, but also DwN, since it lies very close to \ntr\ in the projected phase-space diagram and has an \HI\ tail pointing towards it (see Fig.~\ref{fig:vegasHI}). In summary, the analysis of the locations and motions of the \HI\ detections within the \ntr\ group supports a scenario in which different galaxies recently interacted with the \ntr, and possibly the dark \HI\ cloud is the remnant of one of these interactions, and survived in the inter-galactic medium. 

\begin{table}[tbh]
        \caption{Optical properties of the \HI\ detections in the \ntr\ group.}
        \centering
        \label{tab:stars}
        \begin{tabularx}{\columnwidth}{X c c c c c c} 
                \hline\hline  
Source  &$\mu_g$ [mag] & $\mu_r$ [mag] & $ g-r$ & $M_{\star} (\rm g,r)$    \\
Name	&              &               &        &[$\times10^{10}$\msun] \\
\hline
\ntr & $11.46$ & $10.68$ & 0.78 & 21.06, 17.01 \\
\mbox{ESO~435-028}& $15.53$&$15.22$& 0.31 & 0.069, 0.057 \\
\nzr& $11.81$&$11.28$& 0.53 & 5.56, 4.49 \\
\mbox{EQJ100-3124}&$16.17$&$15.47$& 0.7 & 0.19, 0.16 \\
\mbox{NGC~3108}&$11.80$&$10.99$& 0.81 & 17.02, 13.74 \\
\dwn&$17.00$&$16.71$& 0.29 & 0.017, 0.014 \\
\mbox{DwW}&$15.66$&$15.44$& 0.22 & 0.044,  0.035 \\
\hline
         \end{tabularx}
        \tablefoot{The mean errors on the magnitudes are $0.05$ and $0.04$ for the g and r bands, respectively.}
\end{table}

\begin{figure}[tbh]
	\centering
	\includegraphics[width=\columnwidth]{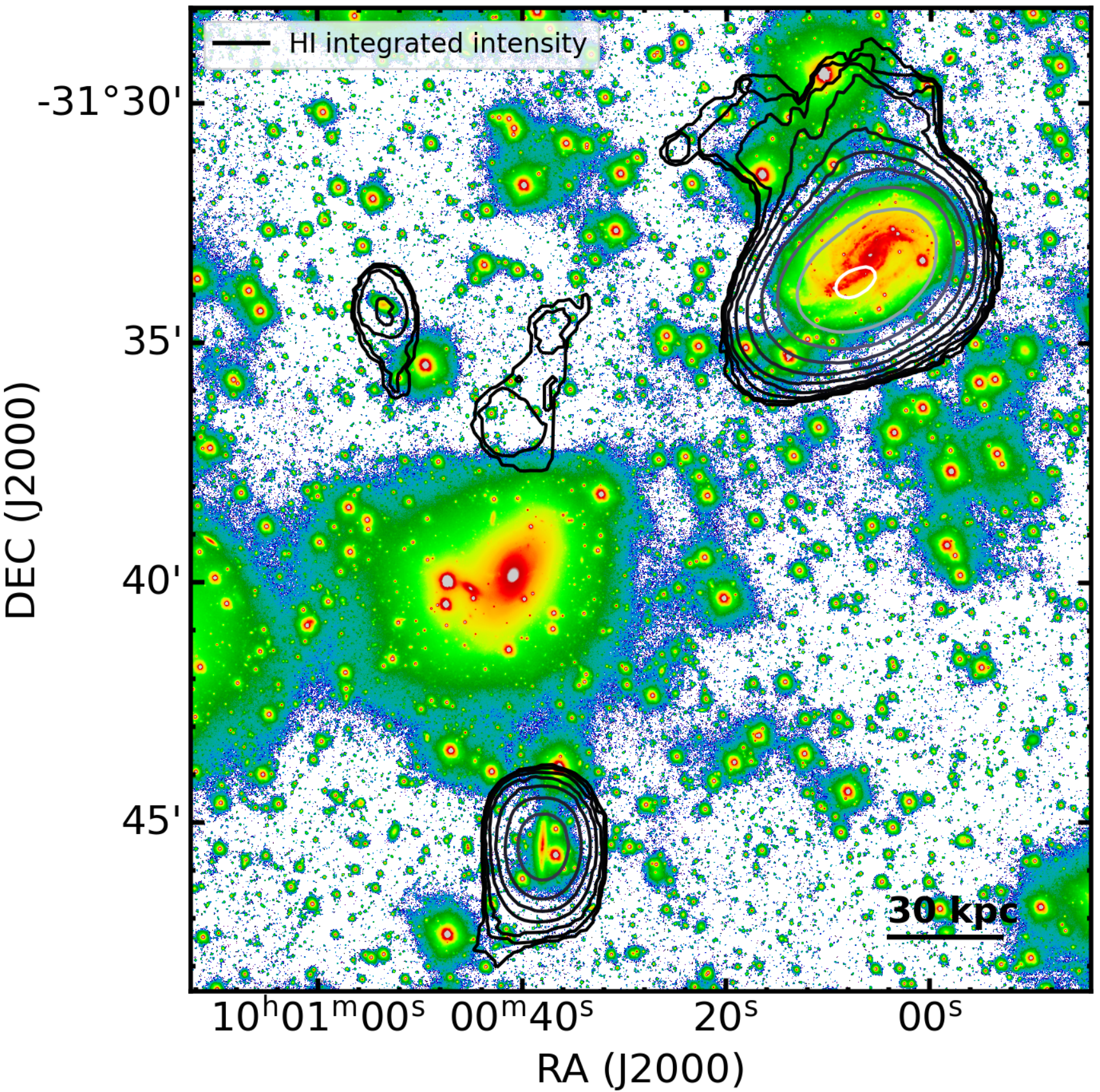}	
	\caption{VEGAS image of the north of \ntr. The deep photometric observations show the absence of a galaxy in correspondence of the \HI\ cloud detected by the ATCA observations. \HI\ contours are as in Fig.~\ref{fig:vegasHI}.}
	\label{fig:cloudOptical}%
\end{figure} 

\begin{figure}[tbh]
	\centering
	\includegraphics[width=\columnwidth]{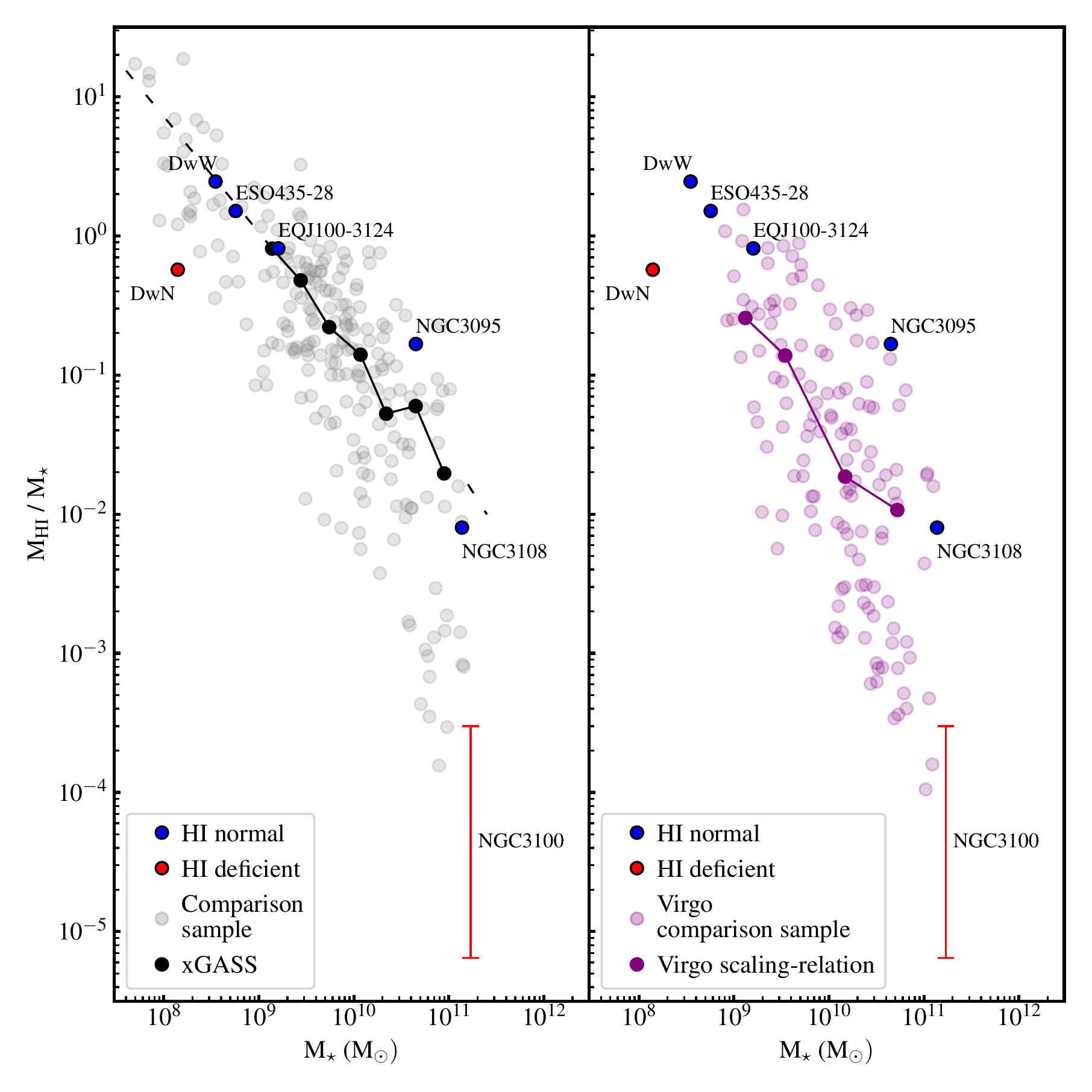}
	\caption{{\em Left Panel}: \HI\ fraction vs. the stellar mass of the galaxies in the \ntr\ group (red+blue markers) compared with non-cluster galaxies from VSG+HRS~\citep[grey circles, ][]{Kreckel:2012,Boselli:2014}. Red and blue colours show \HI\ deficient and normal galaxies, respectively. We show with a black solid line the xGASS scaling relation~\citep{Catinella:2018}. The black dashed line is the linear extrapolation of this trend for M$_\star~<~1.4~\times~10^9$~\msun. {\em Right Panel}: Similar to the left panel. Here we show the comparison between the \HI\ mass fraction of the galaxies in the \ntr\ group with that of Virgo cluster galaxies from HRS (light purple circles). Dark purple circles show the average scaling relation obtained from Virgo cluster galaxies~\citep{Cortese:2016}. }
 	\label{fig:groupHI}%
\end{figure}

Since recent minor interactions likely characterised the assembly history of the centre of the \ntr\ group, we ask ourselves whether these interactions have already removed an important amount of \HI\ from its members. 
One way to answer this question is to compare their \HI\ content with the one of galaxies that live in poorer environments (where sources likely do not experience any significant event of gas removal such as mergers and ram pressure stripping; and galaxies that live in denser environments (e.g.\,clusters, where instead members likely experience several gas removal events). Following \citet{Loni:2021} and \citet{Molnar:2022}, for the comparison with galaxies in poorer environments, we select all the field sources in the Local Universe with \HI\ and stellar contents measured in the Herschel Reference Survey (excluding the Virgo cluster members) and Void Galaxy Survey~\citep[][respectively]{Boselli:2014,Kreckel:2012}. For galaxies in denser environments, we use the sources detected in the Virgo cluster~\citep[][]{Cortese:2016}.
The left panel of Fig.~\ref{fig:groupHI} shows the comparison between the \HI\ mass fraction (defined as the ratio between the \HI\ mass and the stellar mass) versus stellar mass for the \HI\ detections in the \ntr\ group and the galaxies of the comparison sample. The dotted line shows the GALEX Arecibo SDSS Survey~\citep[xGASS, ][]{Catinella:2018} scaling relation, \ie\ the median value of the \HI\ content for galaxies in the Local Group. We extrapolated the xGASS trend down to 10$^7$~M$_\odot$ (black dashed line), which is consistent with the \HI\ mass fraction of HRS+VGS galaxies. Given that for the dark cloud {\em Cl} there is no corresponding stellar counterpart (and thus we cannot estimate a stellar mass), we do not plot this source in Fig.~\ref{fig:groupHI}.

The galaxies of the \ntr\ group do not show any significant offset with respect to those of the comparison samples in the plane of the \HI\ mass fraction vs stellar mass. Overall, it appears that members of this group host an amount of \HI\ consistent with that of galaxies which do not live in dense environments. To quantify any possible case of \HI\ deficiency, we also evaluate the deviation from the xGASS scaling relations for both our \HI\ sources and the galaxies in the comparison sample. We consider as \HI\ deficients all sources of the \ntr\ group whose offset from the scaling relations is larger than $1\sigma$ with respect to the standard deviation of the VGS+HRS sample. These galaxies are shown in red in Fig.~\ref{fig:groupHI}. One of the two sources is \ntr, which is 2--3 orders of magnitude offset with respect to the scaling relation. As illustrated in Sect.~\ref{sec:hiAbs}, however, in this galaxy \HI\ is detected in absorption, which limits an accurate estimate of the total gas mass (which may range between $10^6$ and $10^7$~\msun). Nevertheless, if more \HI\ is present within the stellar body of \ntr, it must have very low column densities for not being detected in emission (\nhi$\lesssim 10^{19}$~\cmsq), and would thus likely not contribute significantly to the total \HI\ mass. 
Moreover, \ntr\ is the most massive ETG of its group, and its \HI\ content (even if a lower limit) is consistent with what expected in massive elliptical galaxies in groups, where often little amount of \HI\ (if any) is detected  \citep[][]{Odekon:2016}.

The other group member resulting \HI\ deficient is \dwn. This further supports (along with the disturbed \HI\ morphology and the indications of the phase-space diagram; see Section~\ref{sec:hiRes}) a scenario in which this source may have undergone a minor interaction with \ntr. The other \HI\ detected dwarf galaxy, \dww, is not deficient in \HI\ and has a regular \HI\ morphology, further indicating that this source did not enter yet in the virialised region of the group.

In the right panel of Fig.~\ref{fig:groupHI} we show the comparison between the \HI\ content in the \ntr\ group and the galaxies of the Virgo cluster. The purple dotted line shows the \HI\ mass fraction scaling relation obtained for Virgo galaxies~\citep[][]{Cortese:2016}. Besides our \HI\ deficient galaxies, all group members are above the Virgo scaling relation. Although a direct comparison with \dwn\ is hard due to the lack of Virgo galaxies below $\sim10^{9}$~\msun, \dwn\ might have reached a level of \HI\ deficiency similar to that of Virgo galaxies. The interaction history of this dwarf satellite may thus be similar to that of galaxies found in much denser environments.

In summary, the analysis of the location and \HI\ morphologies of the sources we detect in the \ntr\ group suggests that minor interactions affect the \HI\ disks of most of them. In particular \nzr\ and \dwn\ have most likely recently interacted with the \ntr\ such that part of their gas has been stripped out, leaving clear \HI\ tails in the outskirts of their disks. The origin of the dark cloud Cl remains puzzling. On the one hand, the cloud is located in the field between \ntr\ and \nzr, but its velocities are blue-shifted with respect to both. On the other hand, \dwn\ is likely \HI\ depleted and the cloud is also closer to it in projection than to~\nzr. Despite the controversial origin of the dark cloud, the \HI\ content and distribution of the \ntr\ group indicate that several sources close to \ntr\ are rich in cold gas and that minor interactions with it may have contributed to replenishing this ETG of cold atomic (and molecular) gas. During the interactions the dark cloud may have acquired a sufficient velocity offset ($\sim 200$\kms) to survive into the IGM and not be re-accreted (recently) onto the sources. 

\subsection{The \HI\ disk of \ntr}
\label{sec:hico}

\begin{figure*}[tbh]
	\centering
    \includegraphics[width=\textwidth]{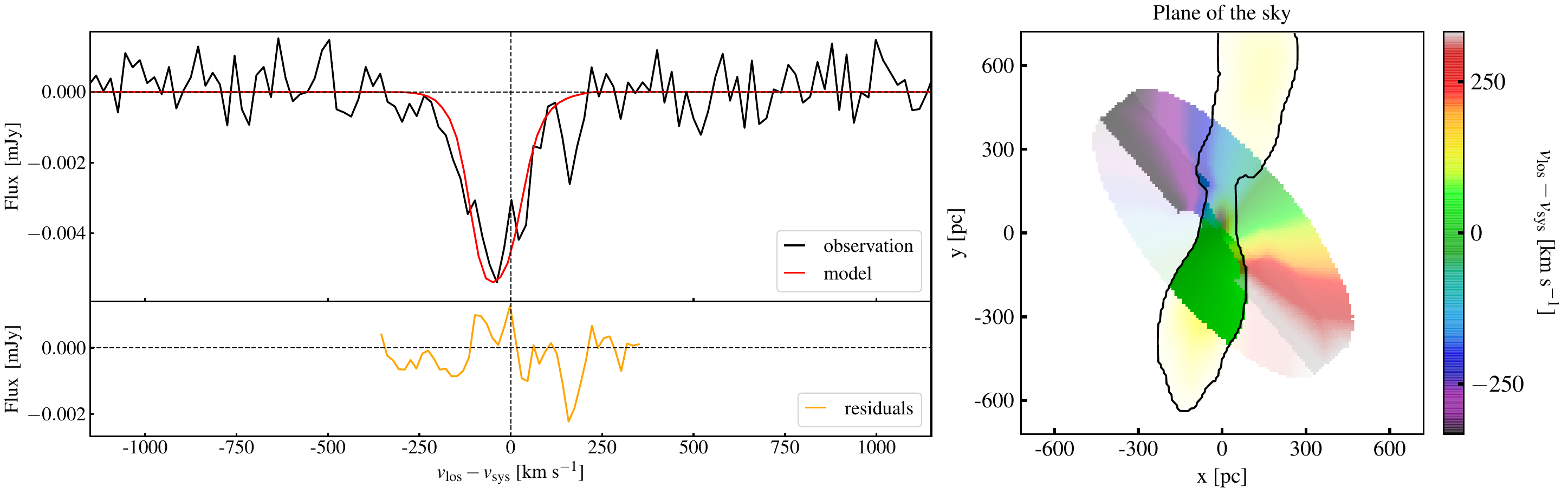}
	\caption{{\em Left Panel}: Best-fit \HI\ absorption generated by a disk in front of the radio continuum, overlaid with the observed line. Residuals are shown in the bottom panel. {\em Right Panel}: Orientation of the \HI\ disk generating the line in the left panel in the plane of the sky. The absorbed part of the disk is against the core and the southern jet and is shown in green. The radio continuum is shown in the background, the colour-scale shows the rotation of the disk relative to systemic. The inclination and position angle of the disk best reproducing the observed \HI\ are the same as the \CO\ disk resolved by ALMA~\citep[see, for example, the right panel of Fig.~\ref{fig:vegasHI} or Fig. 7 of ][]{Ruffa:2019a}}
	\label{fig:absMod}%
\end{figure*}

From the \HI\ absorption complex of \ntr\ and the morphology of its radio source it is possible to infer the overall distribution of the \HI\ in this AGN. As mentioned in Sect.~\ref{sec:hiAbs}, an absorption line peaking close to the systemic velocity such as that detected in NGC\,3100 likely traces a rotating disk. 
The \HI\ detected in absorption may thus be the part against the radio continuum of a diffuse disk extending throughout the entire stellar body of the galaxy, as seen in several nearby ETGs~\citep[][]{Serra:2012,Maccagni:2017}. Alternatively, the \HI\ gas could be concentrated just in the innermost regions of \ntr\ and form a circum-nuclear disk. This is often observed in radio AGN embedded in an S0 galaxy~\citep[see, for example, \mbox{PKS~1718-649}, Fornax A, Centaurus A,][]{Maccagni:2014,Maccagni:2021,Struve:2010,Morganti:2010}. Such circum-nuclear disks are typically multi-phase, with the neutral gas showing kinematics similar to that of the molecular phase~\citep[\eg][]{Maccagni:2016,Maccagni:2018,McCoy:2017,Espada:2017}. As anticipated in Section~\ref{sec:ntr}, a circum-nuclear molecular gas disk has been already detected in NGC\,3100~\citet[][see also the right panel of Fig.~\ref{fig:vegasHI}]{Ruffa:2019a,Ruffa:2019b,Ruffa:2022}. Hence, it is plausible to assume that the neutral and molecular gas phases co-exist also in the circum-nuclear regions of NGC\,3100. As a proof-of-concept, we investigate if and what part of the \HI\ absorption complex traces the circum-nuclear molecular disk. To this aim, we build a toy-model of an \HI\ disk with the same extent, thickness and velocity curve of the \CO\ disk~\citep[$PA=220^\circ$, $i=60^\circ$, $v_{\rm rot} =345$~\kms; see the bottom right panel of Fig.~\ref{fig:cloudOptical} in ][]{Ruffa:2022}. With a procedure illustrated in detail in~\cite{Maccagni:2017b}\footnote{using the open-source software \href{https://github.com/Fil8/MoD_AbS}{\tt MoD\_AbS}}, we overlay the mock \HI\ disk on the high-resolution continuum image of \ntr\ to determine how absorption line complex produced by such disk would look like. Since the radio continuum is unresolved by the ATCA observations, we assume that at 1.4 GHz it has the same morphology and flux distribution than as 5 GHz, where we resolve the radio jets (see right panel of Fig.~\ref{fig:vegasHI}). This enables us to understand if the kinematics of a disk with the same properties of the CO disk would reproduce the kinematics of the \HI, i.e. shape of the absorption line. Since we do not know of the \HI\ disk and we have to make assumptions on the one of the radio continuum we normalize the peak of the modelled absorption line to the observed one. Given that the core is approximately five times brighter than the jets, likely the part of the disk in front if it is the one most contributing to the absorption.
The results are shown in Fig.~\ref{fig:absMod}. The bulk of the \HI\ absorption feature is clearly well reproduced by the modelled disk, where part of it intercepts along the line of sight the core and the radio jet. This indicates the \HI\ absorption likely traces the same circum-nuclear disk seen in emission in the molecular phase, rather than a component at outer distances or a disk with different orientation. In this configuration, about one sixth of the disk is traced by the absorption line. Assuming uniform distribution, the total \HI\ mass of the circum-nuclear disk is thus $1.1\times10^7$~\msun. This value is compatible with the upper limit estimated in Sect.~\ref{sec:hiAbs}.

Given that the total molecular gas mass is likely in the range $0.5-3.5\times10^8$~\msun~\citep[][]{Ruffa:2022}, the \Htwo/\HI\ ratio in \ntr\ is $\sim 1-10$. This value is similar to what found in other nearby LERGs with a dust lane in the centre such as - for example - \mbox{NGC\,1316}~\citep[also known as Fornax A, see e.g.\,][]{Morokuma-Matsui:2019,Maccagni:2021}, and suggests that the bulk of the mass of the circum-nuclear disk is cold and is likely being converted from atomic to molecular. Moreover, since the dust formation and conversion from atomic to molecular gas are enhanced in interacting galaxies~\citep[\eg][]{Sanders:1996,Nakanishi:2006,Kaneko:2017}, it is possible that the small interactions and consequent infall of cold gas is enhancing the conversion to molecular gas.

The nature of the redshifted component of the \HI\ absorption remains puzzling. The modeled circum-nuclear disk shown in Fig.~\ref{fig:absMod} does not reproduce its velocities. The molecular disk shows several clouds with small (within the rotational velocity of the galaxy) deviations from regular rotation, some of which are outflowing while others may be fuelling the AGN~\citep[see][]{Ruffa:2022}. It is possible that the redshifted \HI\ feature traces the neutral phase of some of these clouds. Given its location in front of the radio source and its redshifted velocities this feature may trace \HI\ clouds that are directed towards the central SMBH, thus contributing to its fuelling. Another possibility is that the redshifted feature is tracing \HI\ clouds currently falling onto the disk, as a consequence of the recent minor interactions between \ntr\ and its satellites. Higher sensitivity and resolution observations are required to discriminate between the two possibilities. Such observations would also enable us to connect the \HI\ detected in absorption, with that detected in emission in the outskirts of our target, thus fully determining its origin and role in powering the nuclear activity of this galaxy.

\section{Summary and future prospects}
\label{sec:conclusions}

Deep ATCA 1.4~GHz observations of the \ntr\ loose group show, for the first time, that several satellites of \ntr\ have significant amounts of neutral hydrogen. Most \HI\ disks have asymmetric morphologies, with low-column density tails in their outskirts (Sect.~\ref{sec:hiRes} and Fig.~\ref{fig:vegasHI}). This, along with the projected location of the group members (Sect.~\ref{sec:group} and Fig.~\ref{fig:phaseSpace}), suggests that recent minor interactions have occurred between \ntr\ and its two closest satellites (\nzr\ and \dwn), likely channeling relatively large amounts of cold gas and dust in its centre. These interactions must have been weak to not significantly deplete of gas the satellites, but only unsettle the outer regions of their \HI\ disks, as suggested by the \HI-to-stellar mass ratio of their group members (Fig.~\ref{fig:groupHI}). This scenario is also supported by the deep optical photometry of the group presented in this paper (and acquired as part of the VEGAS project), showing that the stellar bodies of these galaxies are not disturbed and tidal stellar streams are not detected (Fig.~\ref{fig:cloudOptical}).

We detect a diffuse dark cloud between \ntr\ and its companion, \nzr. The combination of the ATCA and VEGAS observations allows us to determine that the cloud does not have a stellar counterpart (see Fig.~\ref{fig:cloudOptical}). This cloud could be the remnant of an interaction between \nzr, or \dwn, and \ntr. 
The dark cloud could also form an \HI\ reservoir in the outskirts of \ntr\ that may replenish its central regions of cold gas. Deeper and higher resolution observations are needed to further investigate our hypotheses.\\
The \HI\ absorption complex detected for the very first time in \ntr\ indicates that the circum-nuclear ISM of this LERG has both a molecular~\citep[\eg ][]{Ruffa:2019b} and a neutral component (Sect.~\ref{sec:hiAbs}). In particular, most of the \HI\ seems to be distributed in the same circum-nuclear disk as the molecular gas (Sect.~\ref{sec:hico}). The redshifted component of the \HI\ absorption line may trace \HI\ clouds that are deviating from the regular kinematics of the disk. These could be the neutral counterpart of the molecular gas clouds that may be fuelling the nuclear activity~\citep[][]{Ruffa:2022}.  Nevertheless, the nature of this component remains puzzling and how the nuclear activity is being sustained remains unclear.

The ATCA observations  of the \mbox{NGC\,3100} group, combined with the new deep optical images from VEGAS,  presented in this paper, allow us to determine that the cold gas at the centre of \ntr\ likely originates from minor interactions underwent with its satellites. Clarifying how the gas is then funnelled onto the circum-nuclear regions and which is the exact mechanism powering the AGN needs deeper and higher resolution \HI\ observations. These would enable us to connect the \HI\ detected in emission in the outskirts of \ntr\ with that detected in absorption in its centre. 
A major step forward in the study of nearby AGN has now been made thanks to the novel radio interferometer MeerKAT~\citep[][]{Camilo:2018,Mauch:2020}. This telescope has a unique combination of long baselines enabling high resolution observations ($\sim 10$\arcsec) over a large field-of-view ($\sim1.5$ deg$^2$). Moreover, the dense core of MeerKAT, along with its 64 antennas, enables high resolution observations of the diffuse \HI\ (\nhi$\sim 10^{19}$\cmsq) which is typically found in the circum-nuclear regions of AGN (as in \ntr). For example, \meer\ L-band observations revealed for the very first time a massive ($2\times 10^8$\msun), low-column density (\nhi$\sim 10^{19}$\cmsq) reservoir of \HI\ in the centre and several filaments in the outskirts of the nearby LERG \mbox{Fornax~A} (\mbox{NGC 1316})~\citep[][]{Serra:2019,Kleiner:2021}. Combining the \HI\ information with the molecular and ionised gas observed in the same central regions of this radio galaxy, it has been possible to reveal the co-existence of AGN feeding and feedback mechanisms involving the multi-phase circum-nuclear ISM and determine how chaotic cold accretion may self-sustain the rapid flickering of the AGN~\citep[][]{Maccagni:2020,Maccagni:2021}.

Up-coming MeerKAT observations of \ntr\ will allow us to further investigate the \HI\ in its group and the nature of its absorption complex. Similarly to Fornax~A, we will be able to combine the \HI\ and CO observations, and put more solid constraints on the feeding mechanism of this LERG and the effects of the expansion of its radio jets. Thanks to the \meer\ sensitivity ($\sim 3\times 10^{18}$\cmsq) and high resolution ($\lesssim 20$\arcsec, $\Delta v\sim 1.4$~\kms), we will likely connect the \HI\ in the centre of \ntr\ with that in the outskirts and in the group, and fully determine the origin of the dark cloud.

\begin{acknowledgements}
This project has received funding from the European Research Council (ERC) under the European Union’s Horizon 2020 research and innovation programme (grant agreement no.~679627 and grant agreement no. 882793). IP acknowledges support from the INAF SKA/CTA PRIN “FORECaST” project and from the PRIN MIUR project “Black Hole winds and the Baryon Life Cycle of Galaxies: the stone-guest at the galaxy evolution supper”, contract 2017PH3WAT. The Australia Telescope Compact Array is part of the Australia Telescope which is funded by the Commonwealth of Australia for operation as a National Facility managed by CSIRO.
\end{acknowledgements}

\bibliographystyle{aa} 
\bibliography{n3100HI.bib} 

\begin{thebibliography}{87}
\expandafter\ifx\csname natexlab\endcsname\relax\def\natexlab#1{#1}\fi

\bibitem[{{Babyk} {et~al.}(2019){Babyk}, {McNamara}, {Tamhane}, {Nulsen},
  {Russell}, \& {Edge}}]{Babyk:2019}
{Babyk}, I.~V., {McNamara}, B.~R., {Tamhane}, P.~D., {et~al.} 2019, \apj, 887,
  149

\bibitem[{{Best} \& {Heckman}(2012)}]{Best:2012}
{Best}, P.~N. \& {Heckman}, T.~M. 2012, \mnras, 421, 1569

\bibitem[{{Boselli} {et~al.}(2014){Boselli}, {Cortese}, \&
  {Boquien}}]{Boselli:2014}
{Boselli}, A., {Cortese}, L., \& {Boquien}, M. 2014, \aap, 564, A65

\bibitem[{{Camilo} {et~al.}(2018){Camilo}, {Scholz}, {Serylak}, {Buchner},
  {Merryfield}, {Kaspi}, {Archibald}, {Bailes}, {Jameson}, {van Straten},
  {Sarkissian}, {Reynolds}, {Johnston}, {Hobbs}, {Abbott}, {Adam}, {Adams},
  {Alberts}, {Andreas}, {Asad}, {Baker}, {Baloyi}, {Bauermeister}, {Baxana},
  {Bennett}, {Bernardi}, {Booisen}, {Booth}, {Botha}, {Boyana}, {Brederode},
  {Burger}, {Cheetham}, {Conradie}, {Conradie}, {Davidson}, {De Bruin}, {de
  Swardt}, {de Villiers}, {de Villiers}, {de Villiers}, {de Villiers}, {De
  Waal}, {Dikgale}, {du Toit}, {du Toit}, {Esterhuyse}, {Fanaroff}, {Fataar},
  {Foley}, {Foster}, {Fourie}, {Gamatham}, {Gatsi}, {Geschke}, {Goedhart},
  {Grobler}, {Gumede}, {Hlakola}, {Hokwana}, {Hoorn}, {Horn}, {Horrell},
  {Hugo}, {Isaacson}, {Jacobs}, {Jansen van Rensburg}, {Jonas}, {Jordaan},
  {Joubert}, {Joubert}, {J{\'o}zsa}, {Julie}, {Julius}, {Kapp}, {Karastergiou},
  {Karels}, {Kariseb}, {Karuppusamy}, {Kasper}, {Knox-Davies}, {Koch},
  {Kotz{\'e}}, {Krebs}, {Kriek}, {Kriel}, {Kusel}, {Lamoor}, {Lehmensiek},
  {Liebenberg}, {Liebenberg}, {Lord}, {Lunsky}, {Mabombo}, {Macdonald},
  {Macfarlane}, {Madisa}, {Mafhungo}, {Magnus}, {Magozore}, {Mahgoub}, {Main},
  {Makhathini}, {Malan}, {Malgas}, {Manley}, {Manzini}, {Marais}, {Marais},
  {Marais}, {Maree}, {Martens}, {Matshawule}, {Matthysen}, {Mauch}, {McNally},
  {Merry}, {Millenaar}, {Mjikelo}, {Mkhabela}, {Mnyandu}, {Moeng}, {Mokone},
  {Monama}, {Montshiwa}, {Moss}, {Mphego}, {New}, {Ngcebetsha}, {Ngoasheng},
  {Niehaus}, {Ntuli}, {Nzama}, {Obies}, {Obrocka}, {Ockards}, {Olyn}, {Oozeer},
  {Otto}, {Padayachee}, {Passmoor}, {Patel}, {Paula}, {Peens-Hough},
  {Pholoholo}, {Prozesky}, {Rakoma}, {Ramaila}, {Rammala}, {Ramudzuli},
  {Rasivhaga}, {Ratcliffe}, {Reader}, {Renil}, {Richter}, {Robyntjies},
  {Rosekrans}, {Rust}, {Salie}, {Sambu}, {Schollar}, {Schwardt}, {Seranyane},
  {Sethosa}, {Sharpe}, {Siebrits}, {Sirothia}, {Slabber}, {Smirnov}, {Smith},
  {Sofeya}, {Songqumase}, {Spann}, {Stappers}, {Steyn}, {Steyn}, {Strong},
  {Struthers}, {Stuart}, {Sunnylall}, {Swart}, {Taljaard}, {Tasse}, {Taylor},
  {Theron}, {Thondikulam}, {Thorat}, {Tiplady}, {Toruvanda}, {van Aardt}, {van
  Balla}, {van den Heever}, {van der Byl}, {van der Merwe}, {van der Merwe},
  {van Niekerk}, {van Rooyen}, {van Staden}, {van Tonder}, {van Wyk}, {Wait},
  {Walker}, {Wallace}, {Welz}, {Williams}, {Xaia}, {Young}, \&
  {Zitha}}]{Camilo:2018}
{Camilo}, F., {Scholz}, P., {Serylak}, M., {et~al.} 2018, \apj, 856, 180

\bibitem[{{Capaccioli} {et~al.}(2015){Capaccioli}, {Spavone}, {Grado},
  {Iodice}, {Limatola}, {Napolitano}, {Cantiello}, {Paolillo}, {Romanowsky},
  {Forbes}, {Puzia}, {Raimondo}, \& {Schipani}}]{capaccioli:2015}
{Capaccioli}, M., {Spavone}, M., {Grado}, A., {et~al.} 2015, \aap, 581, A10

\bibitem[{{Catinella} {et~al.}(2018){Catinella}, {Saintonge}, {Janowiecki},
  {Cortese}, {Dav{\'e}}, {Lemonias}, {Cooper}, {Schiminovich}, {Hummels},
  {Fabello}, {Ger{\'e}b}, {Kilborn}, \& {Wang}}]{Catinella:2018}
{Catinella}, B., {Saintonge}, A., {Janowiecki}, S., {et~al.} 2018, \mnras, 476,
  875

\bibitem[{{Ching} {et~al.}(2017){Ching}, {Croom}, {Sadler}, {Robotham},
  {Brough}, {Baldry}, {Bland-Hawthorn}, {Colless}, {Driver}, {Holwerda},
  {Hopkins}, {Jarvis}, {Johnston}, {Kelvin}, {Liske}, {Loveday}, {Norberg},
  {Pracy}, {Steele}, {Thomas}, \& {Wang}}]{Ching:2017}
{Ching}, J.~H.~Y., {Croom}, S.~M., {Sadler}, E.~M., {et~al.} 2017, \mnras, 469,
  4584

\bibitem[{{Choi} {et~al.}(2015){Choi}, {Ostriker}, {Naab}, {Oser}, \&
  {Moster}}]{Choi:2015}
{Choi}, E., {Ostriker}, J.~P., {Naab}, T., {Oser}, L., \& {Moster}, B.~P. 2015,
  Monthly Notices of the Royal Astronomical Society, 449, 4105

\bibitem[{{Cortese} {et~al.}(2016){Cortese}, {Bekki}, {Boselli}, {Catinella},
  {Ciesla}, {Hughes}, {Baes}, {Bendo}, {Boquien}, {de Looze}, {Smith},
  {Spinoglio}, \& {Viaene}}]{Cortese:2016}
{Cortese}, L., {Bekki}, K., {Boselli}, A., {et~al.} 2016, \mnras, 459, 3574

\bibitem[{{Curran} \& {Duchesne}(2018)}]{Curran:2018}
{Curran}, S.~J. \& {Duchesne}, S.~W. 2018, \mnras, 476, 3580

\bibitem[{{Davis} {et~al.}(2019){Davis}, {Greene}, {Ma}, {Blakeslee}, {Dawson},
  {Pandya}, {Veale}, \& {Zabel}}]{Davis:2019}
{Davis}, T.~A., {Greene}, J.~E., {Ma}, C.-P., {et~al.} 2019, \mnras, 486, 1404

\bibitem[{{D'Onofrio} {et~al.}(2021){D'Onofrio}, {Marziani}, \&
  {Chiosi}}]{Donofrio:2021}
{D'Onofrio}, M., {Marziani}, P., \& {Chiosi}, C. 2021, Frontiers in Astronomy
  and Space Sciences, 8, 157

\bibitem[{{Ekers} {et~al.}(1989){Ekers}, {Wall}, {Shaver}, {Goss}, {Fosbury},
  {Danziger}, {Moorwood}, {Malin}, {Monk}, \& {Ekers}}]{Ekers:1989}
{Ekers}, R.~D., {Wall}, J.~V., {Shaver}, P.~A., {et~al.} 1989, \mnras, 236, 737

\bibitem[{{Emonts} {et~al.}(2008){Emonts}, {Morganti}, {Oosterloo}, {Holt},
  {Tadhunter}, {van der Hulst}, {Ojha}, \& {Sadler}}]{Emonts:2008}
{Emonts}, B.~H.~C., {Morganti}, R., {Oosterloo}, T.~A., {et~al.} 2008, \mnras,
  387, 197

\bibitem[{{Espada} {et~al.}(2017){Espada}, {Matsushita}, {Miura}, {Israel},
  {Neumayer}, {Martin}, {Henkel}, {Izumi}, {Iono}, {Aalto}, {Ott}, {Peck},
  {Quillen}, \& {Kohno}}]{Espada:2017}
{Espada}, D., {Matsushita}, S., {Miura}, R.~E., {et~al.} 2017, \apj, 843, 136

\bibitem[{{Gaspari} {et~al.}(2015){Gaspari}, {Brighenti}, \&
  {Temi}}]{Gaspari:2015}
{Gaspari}, M., {Brighenti}, F., \& {Temi}, P. 2015, \aap, 579, A62

\bibitem[{{Gaspari} {et~al.}(2017){Gaspari}, {Temi}, \&
  {Brighenti}}]{Gaspari:2017}
{Gaspari}, M., {Temi}, P., \& {Brighenti}, F. 2017, \mnras, 466, 677

\bibitem[{{Ger{\'e}b} {et~al.}(2015){Ger{\'e}b}, {Maccagni}, {Morganti}, \&
  {Oosterloo}}]{Gereb:2015}
{Ger{\'e}b}, K., {Maccagni}, F.~M., {Morganti}, R., \& {Oosterloo}, T.~A. 2015,
  \aap, 575, A44

\bibitem[{{Hardcastle}(2018)}]{Hardcastle:2018}
{Hardcastle}, M. 2018, Nature Astronomy, 2, 273

\bibitem[{{Hardcastle} {et~al.}(2007){Hardcastle}, {Evans}, \&
  {Croston}}]{Hardcastle:2007}
{Hardcastle}, M.~J., {Evans}, D.~A., \& {Croston}, J.~H. 2007, \mnras, 376,
  1849

\bibitem[{{Harrison} {et~al.}(2018){Harrison}, {Costa}, {Tadhunter},
  {Fl{\"u}tsch}, {Kakkad}, {Perna}, \& {Vietri}}]{Harrison:2018}
{Harrison}, C.~M., {Costa}, T., {Tadhunter}, C.~N., {et~al.} 2018, Nature
  Astronomy, 2, 198

\bibitem[{{Hau} {et~al.}(2008){Hau}, {Bower}, {Kilborn}, {Forbes}, {Balogh}, \&
  {Oosterloo}}]{hau:2008}
{Hau}, G. K.~T., {Bower}, R.~G., {Kilborn}, V., {et~al.} 2008, \mnras, 385,
  1965

\bibitem[{{Ineson} {et~al.}(2015){Ineson}, {Croston}, {Hardcastle}, {Kraft},
  {Evans}, \& {Jarvis}}]{Ineson:2015}
{Ineson}, J., {Croston}, J.~H., {Hardcastle}, M.~J., {et~al.} 2015, \mnras,
  453, 2682

\bibitem[{{Iodice} {et~al.}(2021){Iodice}, {Spavone}, {Raj}, {Capaccioli},
  {Cantiello}, \& {VEGAS science team}}]{iodice:2021}
{Iodice}, E., {Spavone}, M., {Raj}, M.~A., {et~al.} 2021, arXiv e-prints,
  arXiv:2102.04950

\bibitem[{Jaff{\'e} {et~al.}(2015)Jaff{\'e}, {Smith}, {Candlish}, {Poggianti},
  {Sheen}, \& {Verheijen}}]{Jaffe:2015}
Jaff{\'e}, Y.~L., {Smith}, R., {Candlish}, G.~N., {et~al.} 2015, \mnras, 448,
  1715

\bibitem[{Jaff{\'e} {et~al.}(2016)Jaff{\'e}, {Verheijen}, {Haines}, {Yoon},
  {Cybulski}, {Montero-Casta{\~n}o}, {Smith}, {Chung}, {Deshev},
  {Fern{\'a}ndez}, {van Gorkom}, {Poggianti}, {Yun}, {Finoguenov}, {Smith}, \&
  {Okabe}}]{Jaffe:2016}
Jaff{\'e}, Y.~L., {Verheijen}, M. A.~W., {Haines}, C.~P., {et~al.} 2016,
  \mnras, 461, 1202

\bibitem[{{Kaneko} {et~al.}(2017){Kaneko}, {Kuno}, {Iono}, {Tamura}, {Tosaki},
  {Nakanishi}, \& {Sawada}}]{Kaneko:2017}
{Kaneko}, H., {Kuno}, N., {Iono}, D., {et~al.} 2017, \pasj, 69, 66

\bibitem[{{Kleiner} {et~al.}(2021){Kleiner}, {Serra}, {Maccagni}, {Venhola},
  {Morokuma-Matsui}, {Peletier}, {Iodice}, {Raj}, {de Blok}, {Comrie},
  {J{\'o}zsa}, {Kamphuis}, {Loni}, {Loubser}, {Moln{\'a}r}, {Passmoor},
  {Ramatsoku}, {Sivitilli}, {Smirnov}, {Thorat}, \& {Vitello}}]{Kleiner:2021}
{Kleiner}, D., {Serra}, P., {Maccagni}, F.~M., {et~al.} 2021, \aap, 648, A32

\bibitem[{{Kourkchi} \& {Tully}(2017)}]{kourkchi:2017}
{Kourkchi}, E. \& {Tully}, R.~B. 2017, \apj, 843, 16

\bibitem[{{Kreckel} {et~al.}(2012){Kreckel}, {Platen}, {Arag{\'o}n-Calvo}, {van
  Gorkom}, {van de Weygaert}, {van der Hulst}, \& {Beygu}}]{Kreckel:2012}
{Kreckel}, K., {Platen}, E., {Arag{\'o}n-Calvo}, M.~A., {et~al.} 2012, \aj,
  144, 16

\bibitem[{{Kuijken}(2011)}]{kuijken2011Msngr.146....8K}
{Kuijken}, K. 2011, The Messenger, 146, 8

\bibitem[{{Laurikainen} {et~al.}(2006){Laurikainen}, {Salo}, {Buta}, {Knapen},
  {Speltincx}, \& {Block}}]{Laurikainen:2006}
{Laurikainen}, E., {Salo}, H., {Buta}, R., {et~al.} 2006, \aj, 132, 2634

\bibitem[{{Loni} {et~al.}(2021){Loni}, {Serra}, {Kleiner}, {Cortese},
  {Catinella}, {Koribalski}, {Jarrett}, {Molnar}, {Davis}, {Iodice},
  {Lee-Waddell}, {Loi}, {Maccagni}, {Peletier}, {Popping}, {Ramatsoku},
  {Smith}, \& {Zabel}}]{Loni:2021}
{Loni}, A., {Serra}, P., {Kleiner}, D., {et~al.} 2021, \aap, 648, A31

\bibitem[{{Maccagni}(2017)}]{Maccagni:2017b}
{Maccagni}, F.~M. 2017, PhD thesis, University of Groningen, Netherlands

\bibitem[{{Maccagni} {et~al.}(2017){Maccagni}, {Morganti}, {Oosterloo},
  {Ger{\'e}b}, \& {Maddox}}]{Maccagni:2017}
{Maccagni}, F.~M., {Morganti}, R., {Oosterloo}, T.~A., {Ger{\'e}b}, K., \&
  {Maddox}, N. 2017, \aap, 604, A43

\bibitem[{{Maccagni} {et~al.}(2014){Maccagni}, {Morganti}, {Oosterloo}, \&
  {Mahony}}]{Maccagni:2014}
{Maccagni}, F.~M., {Morganti}, R., {Oosterloo}, T.~A., \& {Mahony}, E.~K. 2014,
  \aap, 571, A67

\bibitem[{{Maccagni} {et~al.}(2018){Maccagni}, {Morganti}, {Oosterloo}, {Oonk},
  \& {Emonts}}]{Maccagni:2018}
{Maccagni}, F.~M., {Morganti}, R., {Oosterloo}, T.~A., {Oonk}, J.~B.~R., \&
  {Emonts}, B.~H.~C. 2018, \aap, 614, A42

\bibitem[{{Maccagni} {et~al.}(2020){Maccagni}, {Murgia}, {Serra}, {Govoni},
  {Morokuma-Matsui}, {Kleiner}, {Buchner}, {J{\'o}zsa}, {Kamphuis},
  {Makhathini}, {Moln{\'a}r}, {Prokhorov}, {Ramaila}, {Ramatsoku}, {Thorat}, \&
  {Smirnov}}]{Maccagni:2020}
{Maccagni}, F.~M., {Murgia}, M., {Serra}, P., {et~al.} 2020, \aap, 634, A9

\bibitem[{{Maccagni} {et~al.}(2016){Maccagni}, {Santoro}, {Morganti},
  {Oosterloo}, {Oonk}, \& {Emonts}}]{Maccagni:2016}
{Maccagni}, F.~M., {Santoro}, F., {Morganti}, R., {et~al.} 2016, \aap, 588, A46

\bibitem[{{Maccagni} {et~al.}(2021){Maccagni}, {Serra}, {Gaspari}, {Kleiner},
  {Morokuma-Matsui}, {Oosterloo}, {Onodera}, {Kamphuis}, {Loi}, {Thorat},
  {Ramatsoku}, {Smirnov}, \& {White}}]{Maccagni:2021}
{Maccagni}, F.~M., {Serra}, P., {Gaspari}, M., {et~al.} 2021, \aap, 656, A45

\bibitem[{{Mauch} {et~al.}(2020){Mauch}, {Cotton}, {Condon}, {Matthews},
  {Abbott}, {Adam}, {Aldera}, {Asad}, {Bauermeister}, {Bennett}, {Bester},
  {Botha}, {Brederode}, {Brits}, {Buchner}, {Burger}, {Camilo}, {Chalmers},
  {Cheetham}, {de Villiers}, {de Villiers}, {Dikgale-Mahlakoana}, {du Toit},
  {Esterhuyse}, {Fadana}, {Fanaroff}, {Fataar}, {February}, {Frank},
  {Gamatham}, {Geyer}, {Goedhart}, {Gounden}, {Gumede}, {Heywood}, {Hlakola},
  {Horrell}, {Hugo}, {Isaacson}, {J{\'o}zsa}, {Jonas}, {Julie}, {Kapp},
  {Kasper}, {Kenyon}, {Kotz{\'e}}, {Kriek}, {Kriel}, {Kusel}, {Lehmensiek},
  {Loots}, {Lord}, {Lunsky}, {Madisa}, {Magnus}, {Main}, {Malan}, {Manley},
  {Marais}, {Martens}, {Merry}, {Millenaar}, {Mnyandu}, {Moeng}, {Mokone},
  {Monama}, {Mphego}, {New}, {Ngcebetsha}, {Ngoasheng}, {Ockards}, {Oozeer},
  {Otto}, {Patel}, {Peens-Hough}, {Perkins}, {Ramaila}, {Ramudzuli}, {Renil},
  {Richter}, {Robyntjies}, {Salie}, {Schollar}, {Schwardt}, {Serylak},
  {Siebrits}, {Sirothia}, {Smirnov}, {Sofeya}, {Stone}, {Taljaard}, {Tasse},
  {Theron}, {Tiplady}, {Toruvanda}, {Twum}, {van Balla}, {van der Byl}, {van
  der Merwe}, {Van Tonder}, {Wallace}, {Welz}, {Williams}, \&
  {Xaia}}]{Mauch:2020}
{Mauch}, T., {Cotton}, W.~D., {Condon}, J.~J., {et~al.} 2020, \apj, 888, 61

\bibitem[{{McCoy} {et~al.}(2017){McCoy}, {Ott}, {Meier}, {Muller}, {Espada},
  {Mart{\'\i}n}, {Israel}, {Henkel}, {Impellizzeri}, {Aalto}, {Edwards},
  {Brunthaler}, {Neumayer}, {Peck}, {van der Werf}, \& {Feain}}]{McCoy:2017}
{McCoy}, M., {Ott}, J., {Meier}, D.~S., {et~al.} 2017, \apj, 851, 76

\bibitem[{{McFarland} {et~al.}(2013){McFarland}, {Verdoes-Kleijn}, {Sikkema},
  {Helmich}, {Boxhoorn}, \& {Valentijn}}]{McFarland2013}
{McFarland}, J.~P., {Verdoes-Kleijn}, G., {Sikkema}, G., {et~al.} 2013,
  Experimental Astronomy, 35, 45

\bibitem[{{Meyer} {et~al.}(2017){Meyer}, {Robotham}, {Obreschkow}, {Westmeier},
  {Duffy}, \& {Staveley-Smith}}]{Meyer:2017}
{Meyer}, M., {Robotham}, A., {Obreschkow}, D., {et~al.} 2017, \pasa, 34, 52

\bibitem[{{Meyer} {et~al.}(2004){Meyer}, {Zwaan}, {Webster}, {Staveley-Smith},
  {Ryan-Weber}, {Drinkwater}, {Barnes}, {Howlett}, {Kilborn}, {Stevens},
  {Waugh}, {Pierce}, {Bhathal}, {de Blok}, {Disney}, {Ekers}, {Freeman},
  {Garcia}, {Gibson}, {Harnett}, {Henning}, {Jerjen}, {Kesteven}, {Knezek},
  {Koribalski}, {Mader}, {Marquarding}, {Minchin}, {O'Brien}, {Oosterloo},
  {Price}, {Putman}, {Ryder}, {Sadler}, {Stewart}, {Stootman}, \&
  {Wright}}]{Meyer:2004}
{Meyer}, M.~J., {Zwaan}, M.~A., {Webster}, R.~L., {et~al.} 2004, \mnras, 350,
  1195

\bibitem[{{Miraghaei} \& {Best}(2017)}]{Miraghaei:2017}
{Miraghaei}, H. \& {Best}, P.~N. 2017, \mnras, 466, 4346

\bibitem[{{Moln{\'a}r} {et~al.}(2022){Moln{\'a}r}, {Serra}, {van der Hulst},
  {Jarrett}, {Boselli}, {Cortese}, {Healy}, {de Blok}, {Cappellari}, {Hess},
  {J{\'o}zsa}, {McDermid}, {Oosterloo}, \& {Verheijen}}]{Molnar:2022}
{Moln{\'a}r}, D.~C., {Serra}, P., {van der Hulst}, T., {et~al.} 2022, \aap,
  659, A94

\bibitem[{{Morganti}(2010)}]{Morganti:2010}
{Morganti}, R. 2010, \pasa, 27, 463

\bibitem[{{Morganti} \& {Oosterloo}(2018)}]{Morganti:2018}
{Morganti}, R. \& {Oosterloo}, T. 2018, \aapr, 26, 4

\bibitem[{{Morganti} {et~al.}(2001){Morganti}, {Oosterloo}, {Tadhunter}, {van
  Moorsel}, {Killeen}, \& {Wills}}]{Morganti:2001}
{Morganti}, R., {Oosterloo}, T.~A., {Tadhunter}, C.~N., {et~al.} 2001, \mnras,
  323, 331

\bibitem[{{Morokuma-Matsui} {et~al.}(2019){Morokuma-Matsui}, {Serra},
  {Maccagni}, {For}, {Wang}, {Bekki}, {Morokuma}, {Egusa}, {Espada}, {Miura},
  {Nakanishi}, {Koribalski}, \& {Takeuchi}}]{Morokuma-Matsui:2019}
{Morokuma-Matsui}, K., {Serra}, P., {Maccagni}, F.~M., {et~al.} 2019, \pasj,
  71, 85

\bibitem[{{Nagai} {et~al.}(2019){Nagai}, {Onishi}, {Kawakatu}, {Fujita},
  {Kino}, {Fukazawa}, {Lim}, {Forman}, {Vrtilek}, {Nakanishi}, {Noda}, {Asada},
  {Wajima}, {Ohyama}, \& {David}}]{Nagai:2019}
{Nagai}, H., {Onishi}, K., {Kawakatu}, N., {et~al.} 2019, arXiv e-prints,
  arXiv:1905.06017

\bibitem[{{Nakanishi} {et~al.}(2006){Nakanishi}, {Kuno}, {Sofue}, {Sato},
  {Nakai}, {Shioya}, {Tosaki}, {Onodera}, {Sorai}, {Egusa}, \&
  {Hirota}}]{Nakanishi:2006}
{Nakanishi}, H., {Kuno}, N., {Sofue}, Y., {et~al.} 2006, \apj, 651, 804

\bibitem[{{Negri} {et~al.}(2014){Negri}, {Posacki}, {Pellegrini}, \&
  {Ciotti}}]{Negri:2014}
{Negri}, A., {Posacki}, S., {Pellegrini}, S., \& {Ciotti}, L. 2014, \mnras,
  445, 1351

\bibitem[{{Oca{\~n}a Flaquer} {et~al.}(2010){Oca{\~n}a Flaquer}, {Leon},
  {Combes}, \& {Lim}}]{Ocana:2010}
{Oca{\~n}a Flaquer}, B., {Leon}, S., {Combes}, F., \& {Lim}, J. 2010, \aap,
  518, A9

\bibitem[{{O'Dea} \& {Saikia}(2021)}]{ODea:2021}
{O'Dea}, C.~P. \& {Saikia}, D.~J. 2021, \aapr, 29, 3

\bibitem[{{Odekon} {et~al.}(2016){Odekon}, {Koopmann}, {Haynes}, {Finn},
  {McGowan}, {Micula}, {Reed}, {Giovanelli}, \& {Hallenbeck}}]{Odekon:2016}
{Odekon}, M.~C., {Koopmann}, R.~A., {Haynes}, M.~P., {et~al.} 2016, \apj, 824,
  110

\bibitem[{{Oosterloo} {et~al.}(2010){Oosterloo}, {Morganti}, {Crocker},
  {J{\"u}tte}, {Cappellari}, {de Zeeuw}, {Krajnovi{\'c}}, {McDermid},
  {Kuntschner}, {Sarzi}, \& {Weijmans}}]{Oosterloo:2010}
{Oosterloo}, T., {Morganti}, R., {Crocker}, A., {et~al.} 2010, \mnras, 409, 500

\bibitem[{{Oosterloo} {et~al.}(2002){Oosterloo}, {Morganti}, {Sadler},
  {Vergani}, \& {Caldwell}}]{Oosterloo:2002}
{Oosterloo}, T.~A., {Morganti}, R., {Sadler}, E.~M., {Vergani}, D., \&
  {Caldwell}, N. 2002, \aj, 123, 729

\bibitem[{{Poggianti} {et~al.}(2017){Poggianti}, {Jaff{\'e}}, {Moretti},
  {Gullieuszik}, {Radovich}, {Tonnesen}, {Fritz}, {Bettoni}, {Vulcani},
  {Fasano}, {Bellhouse}, {Hau}, \& {Omizzolo}}]{Poggianti:2017}
{Poggianti}, B.~M., {Jaff{\'e}}, Y.~L., {Moretti}, A., {et~al.} 2017, \nat,
  548, 304

\bibitem[{{Prandoni} {et~al.}(2010){Prandoni}, {Laing}, {de Ruiter}, \&
  {Parma}}]{Prandoni:2010}
{Prandoni}, I., {Laing}, R.~A., {de Ruiter}, H.~R., \& {Parma}, P. 2010, \aap,
  523, A38

\bibitem[{{Pulido} {et~al.}(2018){Pulido}, {McNamara}, {Edge}, {Hogan},
  {Vantyghem}, {Russell}, {Nulsen}, {Babyk}, \& {Salom{\'e}}}]{Pulido:2018}
{Pulido}, F.~A., {McNamara}, B.~R., {Edge}, A.~C., {et~al.} 2018, \apj, 853,
  177

\bibitem[{{Ragusa} {et~al.}(2023){Ragusa}, {Iodice}, {Spavone}, {Montes},
  {Forbes}, {Brough}, {Mirabile}, {Cantiello}, {Paolillo}, \&
  {Schipani}}]{Ragusa2023}
{Ragusa}, R., {Iodice}, E., {Spavone}, M., {et~al.} 2023, \aap, 670, L20

\bibitem[{{Ramos Almeida} {et~al.}(2012){Ramos Almeida}, {Bessiere},
  {Tadhunter}, {P{\'e}rez-Gonz{\'a}lez}, {Barro}, {Inskip}, {Morganti}, {Holt},
  \& {Dicken}}]{Almeida:2012}
{Ramos Almeida}, C., {Bessiere}, P.~S., {Tadhunter}, C.~N., {et~al.} 2012,
  \mnras, 419, 687

\bibitem[{{Rhee} {et~al.}(2017){Rhee}, {Smith}, {Choi}, {Yi}, {Jaff{\'e}},
  {Candlish}, \& {S{\'a}nchez-J{\'a}nssen}}]{Rhee:2017}
{Rhee}, J., {Smith}, R., {Choi}, H., {et~al.} 2017, \apj, 843, 128

\bibitem[{{Ruffa} {et~al.}(2019{\natexlab{a}}){Ruffa}, {Davis}, {Prandoni},
  {Laing}, {Paladino}, {Parma}, {de Ruiter}, {Casasola}, {Bureau}, \&
  {Warren}}]{Ruffa:2019b}
{Ruffa}, I., {Davis}, T.~A., {Prandoni}, I., {et~al.} 2019{\natexlab{a}},
  \mnras, 489, 3739

\bibitem[{{Ruffa} {et~al.}(2020){Ruffa}, {Laing}, {Prandoni}, {Paladino},
  {Parma}, {Davis}, \& {Bureau}}]{Ruffa:2020}
{Ruffa}, I., {Laing}, R.~A., {Prandoni}, I., {et~al.} 2020, \mnras, 499, 5719

\bibitem[{{Ruffa} {et~al.}(2022){Ruffa}, {Prandoni}, {Davis}, {Laing},
  {Paladino}, {Casasola}, {Parma}, \& {Bureau}}]{Ruffa:2022}
{Ruffa}, I., {Prandoni}, I., {Davis}, T.~A., {et~al.} 2022, \mnras, 510, 4485

\bibitem[{{Ruffa} {et~al.}(2019{\natexlab{b}}){Ruffa}, {Prandoni}, {Laing},
  {Paladino}, {Parma}, {de Ruiter}, {Mignano}, {Davis}, {Bureau}, \&
  {Warren}}]{Ruffa:2019a}
{Ruffa}, I., {Prandoni}, I., {Laing}, R.~A., {et~al.} 2019{\natexlab{b}},
  \mnras, 484, 4239

\bibitem[{{Sabater} {et~al.}(2015){Sabater}, {Best}, \&
  {Heckman}}]{Sabater:2015}
{Sabater}, J., {Best}, P.~N., \& {Heckman}, T.~M. 2015, \mnras, 447, 110

\bibitem[{{Sanders} \& {Mirabel}(1996)}]{Sanders:1996}
{Sanders}, D.~B. \& {Mirabel}, I.~F. 1996, \araa, 34, 749

\bibitem[{{Sault} {et~al.}(2011){Sault}, {Teuben}, \& {Wright}}]{Sault:2011}
{Sault}, R.~J., {Teuben}, P., \& {Wright}, M. C.~H. 2011, {MIRIAD:
  Multi-channel Image Reconstruction, Image Analysis, and Display}

\bibitem[{{Sault} {et~al.}(1995){Sault}, {Teuben}, \& {Wright}}]{Sault:1995}
{Sault}, R.~J., {Teuben}, P.~J., \& {Wright}, M.~C.~H. 1995, in Astronomical
  Society of the Pacific Conference Series, Vol.~77, Astronomical Data Analysis
  Software and Systems IV, ed. R.~A. {Shaw}, H.~E. {Payne}, \& J.~J.~E.
  {Hayes}, 433

\bibitem[{{Serra} {et~al.}(2019){Serra}, {Maccagni}, {Kleiner}, {de Blok}, {van
  Gorkom}, {Hugo}, {Iodice}, {J{\'o}zsa}, {Kamphuis}, {Kraan-Korteweg}, {Loni},
  {Makhathini}, {Moln{\'a}r}, {Oosterloo}, {Peletier}, {Ramaila}, {Ramatsoku},
  {Smirnov}, {Smith}, {Spavone}, {Thorat}, {Trager}, \& {Venhola}}]{Serra:2019}
{Serra}, P., {Maccagni}, F.~M., {Kleiner}, D., {et~al.} 2019, \aap, 628, A122

\bibitem[{{Serra} {et~al.}(2012){Serra}, {Oosterloo}, {Morganti}, {Alatalo},
  {Blitz}, {Bois}, {Bournaud}, {Bureau}, {Cappellari}, {Crocker}, {Davies},
  {Davis}, {de Zeeuw}, {Duc}, {Emsellem}, {Khochfar}, {Krajnovi{\'c}},
  {Kuntschner}, {Lablanche}, {McDermid}, {Naab}, {Sarzi}, {Scott}, {Trager},
  {Weijmans}, \& {Young}}]{Serra:2012}
{Serra}, P., {Oosterloo}, T., {Morganti}, R., {et~al.} 2012, \mnras, 422, 1835

\bibitem[{{Serra} {et~al.}(2008){Serra}, {Trager}, {Oosterloo}, \&
  {Morganti}}]{serra:2008}
{Serra}, P., {Trager}, S.~C., {Oosterloo}, T.~A., \& {Morganti}, R. 2008, \aap,
  483, 57

\bibitem[{{Serra} {et~al.}(2015){Serra}, {Westmeier}, {Giese}, {Jurek},
  {Fl{\"o}er}, {Popping}, {Winkel}, {van der Hulst}, {Meyer}, {Koribalski},
  {Staveley-Smith}, \& {Courtois}}]{Serra:2015}
{Serra}, P., {Westmeier}, T., {Giese}, N., {et~al.} 2015, \mnras, 448, 1922

\bibitem[{{Smith} {et~al.}(2015){Smith}, {S{\'a}nchez-Janssen}, {Beasley},
  {Candlish}, {Gibson}, {Puzia}, {Janz}, {Knebe}, {Aguerri}, {Lisker},
  {Hensler}, {Fellhauer}, {Ferrarese}, \& {Yi}}]{Smith:2015}
{Smith}, R., {S{\'a}nchez-Janssen}, R., {Beasley}, M.~A., {et~al.} 2015,
  \mnras, 454, 2502

\bibitem[{{Struve} {et~al.}(2010){Struve}, {Oosterloo}, {Morganti}, \&
  {Saripalli}}]{Struve:2010}
{Struve}, C., {Oosterloo}, T.~A., {Morganti}, R., \& {Saripalli}, L. 2010,
  \aap, 515, A67

\bibitem[{{Tamhane} {et~al.}(2022){Tamhane}, {McNamara}, {Russell}, {Edge},
  {Fabian}, {Nulsen}, \& {Babyk}}]{Tamhane:2022}
{Tamhane}, P.~D., {McNamara}, B.~R., {Russell}, H.~R., {et~al.} 2022, \mnras,
  516, 861

\bibitem[{{Temi} {et~al.}(2022){Temi}, {Gaspari}, {Brighenti}, {Werner},
  {Grossova}, {Gitti}, {Sun}, {Amblard}, \& {Simionescu}}]{Temi:2022}
{Temi}, P., {Gaspari}, M., {Brighenti}, F., {et~al.} 2022, \apj, 928, 150

\bibitem[{{Theureau} {et~al.}(1998){Theureau}, {Bottinelli}, {Coudreau-Durand},
  {Gouguenheim}, {Hallet}, {Loulergue}, {Paturel}, \&
  {Teerikorpi}}]{Theureau:1998}
{Theureau}, G., {Bottinelli}, L., {Coudreau-Durand}, N., {et~al.} 1998, \aaps,
  130, 333

\bibitem[{{Thomas} {et~al.}(2021){Thomas}, {Dav{\'e}}, {Jarvis}, \&
  {Angl{\'e}s-Alc{\'a}zar}}]{Thomas:2021}
{Thomas}, N., {Dav{\'e}}, R., {Jarvis}, M.~J., \& {Angl{\'e}s-Alc{\'a}zar}, D.
  2021, \mnras, 503, 3492

\bibitem[{{Tremblay} {et~al.}(2018){Tremblay}, {Combes}, {Oonk}, {Russell},
  {McDonald}, {Gaspari}, {Husemann}, {Nulsen}, {McNamara}, {Hamer}, {O'Dea},
  {Baum}, {Davis}, {Donahue}, {Voit}, {Edge}, {Blanton}, {Bremer}, {Bulbul},
  {Clarke}, {David}, {Edwards}, {Eggerman}, {Fabian}, {Forman}, {Jones},
  {Kerman}, {Kraft}, {Li}, {Powell}, {Randall}, {Salom{\'e}}, {Simionescu},
  {Su}, {Sun}, {Urry}, {Vantyghem}, {Wilkes}, \& {ZuHone}}]{Tremblay:2018}
{Tremblay}, G.~R., {Combes}, F., {Oonk}, J.~B.~R., {et~al.} 2018, \apj, 865, 13

\bibitem[{{Venhola} {et~al.}(2018){Venhola}, {Peletier}, {Laurikainen}, {Salo},
  {Iodice}, {Mieske}, {Hilker}, {Wittmann}, {Lisker}, \&
  {Paolillo}}]{Venhola2018}
{Venhola}, A., {Peletier}, R., {Laurikainen}, E., {et~al.} 2018, A\&A, 620,
  A165

\bibitem[{{Venhola} {et~al.}(2017){Venhola}, {Peletier}, {Laurikainen}, {Salo},
  {Lisker}, {Iodice}, {Capaccioli}, {Verdois Kleijn}, {Valentijn}, {Mieske},
  {Hilker}, {Wittmann}, {van de Ven}, {Grado}, {Spavone}, {Cantiello},
  {Napolitano}, {Paolillo}, \& {Falc{\'o}n-Barroso}}]{Venhola_2017}
{Venhola}, A., {Peletier}, R., {Laurikainen}, E., {et~al.} 2017, A\&A, 608,
  A142

\bibitem[{{Zwaan} {et~al.}(2004){Zwaan}, {Meyer}, {Webster}, {Staveley-Smith},
  {Drinkwater}, {Barnes}, {Bhathal}, {de Blok}, {Disney}, {Ekers}, {Freeman},
  {Garcia}, {Gibson}, {Harnett}, {Henning}, {Howlett}, {Jerjen}, {Kesteven},
  {Kilborn}, {Knezek}, {Koribalski}, {Mader}, {Marquarding}, {Minchin},
  {O'Brien}, {Oosterloo}, {Pierce}, {Price}, {Putman}, {Ryan-Weber}, {Ryder},
  {Sadler}, {Stevens}, {Stewart}, {Stootman}, {Waugh}, \&
  {Wright}}]{Zwaan:2004}
{Zwaan}, M.~A., {Meyer}, M.~J., {Webster}, R.~L., {et~al.} 2004, \mnras, 350,
  1210

\end{thebibliography}

\end{document}